\documentclass[twoside,twocolumn,english,prl]{revtex4-1}

\usepackage[T1]{fontenc}
\usepackage[latin9]{inputenc}
\usepackage{color}
\usepackage{amsmath}
\usepackage{amssymb}
\usepackage{graphicx}
\usepackage{wasysym}

\usepackage[unicode=true,pdfusetitle,
 bookmarks=true,bookmarksnumbered=false,bookmarksopen=false,
 breaklinks=false,pdfborder={0 0 0},pdfborderstyle={},backref=false,colorlinks=true]
 {hyperref}
\hypersetup{
 allcolors=blue}

\makeatletter
\usepackage{braket}
\usepackage{dsfont}

\makeatother

\begin{document}
\title{Experimental detection of microscopic environments using thermodynamic
observables}
\author{Ivan Henao$^{1}$}
\email{ivan.henao@mail.huji.ac.il}

\author{Nadav Katz$^{2}$}
\email{katzn@phys.huji.ac.il}

\author{Raam Uzdin$^{1}$}
\email{raam@mail.huji.ac.il}

\affiliation{$^{1}$Fritz Haber Research Center for Molecular Dynamics,Institute
of Chemistry, The Hebrew University of Jerusalem, Jerusalem 9190401,
Israel}
\affiliation{$^{2}$Racah Institute of Physics, The Hebrew University of Jerusalem,
Jerusalem 9190401, Israel}
\begin{abstract}
Modern thermodynamic theories can be used to study highly complex
quantum dynamics. Here, we experimentally demonstrate that the violation
of thermodynamic constraints allows to detect the coupling of a quantum
system to a hidden environment. By using the IBM quantum superconducting
processors, we perform thermodynamic tests to detect a qubit environment
interacting with a system composed of up to four qubits. The experiments
are complemented by theoretical findings that show efficient scalability
of the tests with respect to system size. Hence, they may be useful
to detect an open system dynamics in situations where other methods
(e.g. quantum state tomography) are practically infeasible. 
\end{abstract}
\maketitle
In recent years various thermodynamic theoretical frameworks (frameworks,
hereafter) for microscopic and/or quantum systems have been formulated
and investigated. Apart from the more traditional approach based on
master equations for open quantum systems \citep{Breuer2002,Kosloff2013},
these frameworks include stochastic thermodynamics \citep{Seifert2012,Sekimoto2010,Esposito2009,Campisi2011},
thermodynamic resource theory \citep{Gour2015,Brandao2015,Horodecki2013,Lostaglio2015,Lostaglio2019},
and several proposals that strongly emphasize the connection between
thermodynamics and information theory \citep{Strasberg2017,Esposito2010,Goold2016,sagawa2013,bera2019thermodynamics,strasberg2020entropy,vinjanampathy2016quantum,benenti2017fundamental,uzdin2018,merali2017new}.
Quantum thermodynamics binds together the thermodynamic frameworks
that are consistent with quantum dynamics \citep{Goold2016,vinjanampathy2016quantum,benenti2017fundamental,merali2017new}. 

Much like the standard Clausius formulation of the second law, each
thermodynamic framework sets constraints on the transformations that
a physical system can undergo. These constraints are to a great extent
determined by the the dynamical protocols that characterize a given
framework. For example, the application of external drivings is allowed
in the derivation of constraints such as fluctuation relations \citep{Esposito2009,Campisi2011}
and thermodynamic uncertainty relations (TURs) \citep{barato2015thermodynamic,macieszczak2018unified,timpanaro2019thermodynamic},
but forbidden in the resource theory approach \citep{Horodecki2013,Lostaglio2019}.
Consequently, a thermodynamic constraint may be violated in a system
whose evolution cannot be described within the corresponding thermodynamic
framework. Here, we apply this ``information from violation'' principle
to evaluate the behavior of devices whose correct operation demands
sufficient isolation from external environments. Emerging quantum
technologies such as quantum computers and quantum simulators fall
within this category. By employing the Melbourne and Essex processors
available through the IBM Quantum Experience (IBMQE) platform, we
experimentally show that the violation of different thermodynamic
constraints can diagnose a non-unitary evolution. 

Fluctuation relations and TURs have been applied to study the operation
of quantum annealers in the D-Wave machine \citep{gardas2018quantum,buffoni2020thermodynamics}.
Fluctuation relations such as that employed in Ref. \citep{gardas2018quantum}
rely on a ``two-point measurement'' scheme, where an initial observable
is measured, the resulting eigenstate is evolved, and at the end of
the evolution another observable is measured \citep{Campisi2011}.
In this case, the experimental verification of the relation (or its
violation, if the device is prone to noise \citep{gardas2018quantum})
involves a sampling of individual trajectories connecting initial
and final eigenstates of the chosen observables. The thermodynamic
tests presented here are constructed by estimating the initial and
final mean values of a suitable observable \citep{uzdin2018global,uzdin2019passivity}.
Thus, in constrast to fluctuation relations, trajectory information
is not required, which can significantly reduce the associated experimental
cost. In particular, we apply recent theoretic-information results
\citep{huang2020predicting} to show that accurate evaluation of the
tests can be achieved with a number of measurements that scales polynomially
in the size of the system. Moreover, their diagnostics capability
is self-contained, meaning that no additional information besides
measurements on the system is necessary (see Fig. 1(a)). This can
be particularly useful in large quantum devices, where error diagnostics
based on the comparison with classical simulation becomes intractable
\citep{arute2019quantum,boixo2018characterizing,zhong2020quantum}. 

Consider a multipartite system prepared in the product of thermal
states 
\begin{equation}
\rho_{s}=\otimes_{i=1}^{n}\frac{e^{-\beta_{i}H_{i}}}{\textrm{Tr}\left(e^{-\beta_{i}H_{i}}\right)},\label{eq:1}
\end{equation}
where $H_{i}$ and $\beta_{i}$ denote respectively the Hamiltonian
and inverse temperature of the $i$th subsystem. For any unital evolution,
the final state $\rho'_{s}$ can be written as $\rho'_{s}=\sum_{i}q_{i}U_{s}^{(i)}(\rho_{s})U_{s}^{(i)\dagger}$
\citep{nielsen2002introduction}, where $\{q_{i}\}$ are probabilities
and $\{U_{s}^{(i)}\}$ are unitary operations. In such a case, the
change in the mean value of the observable $\mathcal{B}\equiv-\textrm{ln}(\rho_{s})=\sum_{i}\beta_{i}H_{i}$
satisfies 
\begin{equation}
\Delta\left\langle \mathcal{B}\right\rangle =\sum_{i=1}^{n}\beta_{i}\Delta\left\langle H_{i}\right\rangle \geq0,\label{eq:2}
\end{equation}
where $\Delta\left\langle H_{i}\right\rangle =\textrm{Tr}[H_{i}(\sum_{i}q_{i}U_{s}^{(i)}\rho_{s}U_{s}^{(i)\dagger}-\rho_{s})]$. 

Equation (\ref{eq:2}) is a reestatement of the Clausius inequality
\citep{lindblad2001non} for initial states (\ref{eq:1}), using the
observable $\mathcal{B}$. A violation of this inequality indicates
that the transformation $\rho_{s}\rightarrow\rho'_{s}$ is non unital
(i.e. it cannot be written as $\rho'_{s}=\sum_{i}q_{i}U_{s}^{(i)}(\rho_{s})U_{s}^{(i)\dagger}$),
which we also describe as a ``heat leak'' \citep{uzdin2018global}.
Since a unitary transformation is a particular case of unital transformation,
the violation of Eq. (\ref{eq:2}) also implies that the system does
not evolve unitarily. The global passivity constraints \citep{uzdin2018global}
are also valid for arbitrary unitary dynamics on initial states of
the form (\ref{eq:1}), and consequently are also useful for heat
leak detection. These constraints rely on an infinite set of ``passive
observables'' $\mathcal{F}$ that satisfy \textcolor{white}{a}
\begin{enumerate}
\item $[\mathcal{F},\mathcal{B}]=0$,
\item The eigenvalues of $\mathcal{F}$ are obtained by applying a non-decreasing
function $f$ to the eigenvalues of $\mathcal{B}$.
\end{enumerate}
The properties 1 and 2 imply that 
\begin{equation}
\Delta\bigl\langle\mathcal{F}\bigr\rangle=\textrm{Tr}[\mathcal{F}(\sum_{i}q_{i}U_{s}^{(i)}\rho_{s}U_{s}^{(i)\dagger}-\rho_{s})]\geq0.\label{eq:3}
\end{equation}
Here, we consider the set of passive observables $\{\mathcal{F}_{\alpha,\delta}\}\equiv\{(\mathcal{B}-\delta\mathbb{I})^{\alpha}\}$,
where $\alpha$ is a positive integer and $\mathbb{I}$ is the identity
operator. Denoting the eigenvalues of $\mathcal{B}$ as $\mathcal{B}_{i}$,
with $\mathcal{B}_{i}\leq\mathcal{B}_{i+1}$, the property 2 restricts
the shift $\delta$ to real values such that $(\mathcal{B}_{i}-\delta)^{\alpha}\leq(\mathcal{B}_{i+1}-\delta)^{\alpha}$
for all $i$. 

It has recently been shown that the mean values of a large number
of observables can be simultaneously estimated in an efficient manner
\citep{huang2020predicting}. That is, with a number of measurements
that does not exhibit an exponential scaing with respect to system
size. Based on this result, we demonstrate in \citep{suppmaterial}
that the number of measurements $N$ required to evaluate the full
family of heat leak tests $\left\{ \Delta\bigl\langle\mathcal{F}_{\alpha,\delta}\bigr\rangle\right\} _{0\leq\alpha\leq\alpha_{\textrm{max}},0\leq\delta\leq\delta_{\textrm{max}}}$
satisfies $N\sim\mathcal{O}\left(\textrm{log}(n)(n+\delta_{\textrm{max}})^{2\alpha_{\textrm{max}}}\right)$.
This polynomial scaling constitutes a potential advantage for the
diagnostics of non-unital errors in large systems, where the application
of more direct methods such as quantum state tomography is infeasible
\citep{haah2017sample}. We theoretically corroborate this advantage
in \citep{suppmaterial}, by deriving an infinite family of heat leaks
that by construction are detectable with any passive observable. The
experiments presented in the following are not intended to provide
experimental evidence of this benefit. Instead, they illustrate the
functionality of the method in a small quantum processor, and show
that passive observables $\mathcal{F}_{\alpha,\delta}$ can outperform
heat leak detection using the Clausius-Lindblad inequality. 

\begin{figure}
\centering{}\includegraphics[scale=0.65]{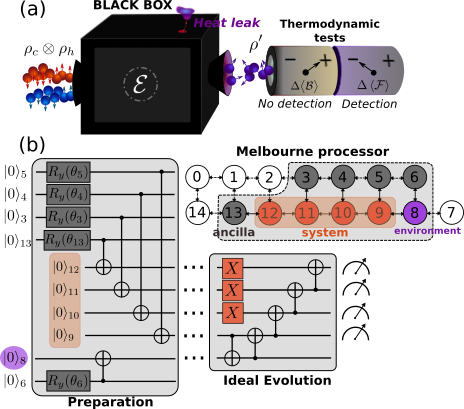}\caption{(a) Thermodynamic tests based on passive observables $\mathcal{F}$
operate as black box tests. This means that they provide unambiguous
heat leak detection without knowing any detail on the dynamics, represented
as a generic CPTP map $\mathcal{E}$. (b) The experiments implemented
with the Melbourne processor involve ten qubits. For the preparation
stage $R_{y}$ rotations by angles $\theta<\pi/2$ are applied to
the qubits 3-6 and 13, which are then entangled with the system qubits
(9-12) and the environment (qubit 8). After this circuit the system
and the environment are left in states $\rho_{s}\sim\rho_{9}\otimes\rho_{10}\otimes\rho_{11}\otimes\rho_{12}$
and $\rho_{e}=\rho_{8}$, where $\rho_{8\protect\leq i\protect\leq12}$
are thermal states at inverse temperatures $\beta_{i}$. The circuit
for the evolution stage contains internal system gates and two cnots
with the environment that generate the heat leak. In practice, this
evolution results in a non-ideal transformation $\rho'_{s}=\tilde{\mathcal{E}}_{s}(\rho_{s})$,
that includes also internal errors. }
\end{figure}

Our experiments involve a preparation stage and an evolution stage.
In the preparation stage the total system is initialized in the state
$\rho_{se}=\rho_{s}\otimes\rho_{e}$, where $\rho_{s}=\otimes_{i}\frac{e^{-\beta_{i}H_{i}}}{Z_{i}}$,
and $\rho_{e}=\frac{e^{-\beta_{e}H_{e}}}{Z_{e}}$ is a thermal state
of the environment at inverse temperature $\beta_{e}$. In the evolution
stage a global unitary $U_{se}$ is applied on $\rho_{se}$, which
includes a system-environment interaction aimed to induce a heat leak.
By measuring the final system state $\rho'_{s}$ in the energy basis,
the heat leak is observed if a violation $\textrm{Tr}[\mathcal{F}(\rho'_{s}-\rho_{s})]<0$
occurs for at least one passive observable $\mathcal{F}$.\textcolor{green}{{}
}
\begin{figure}
\centering{}\includegraphics[scale=0.71]{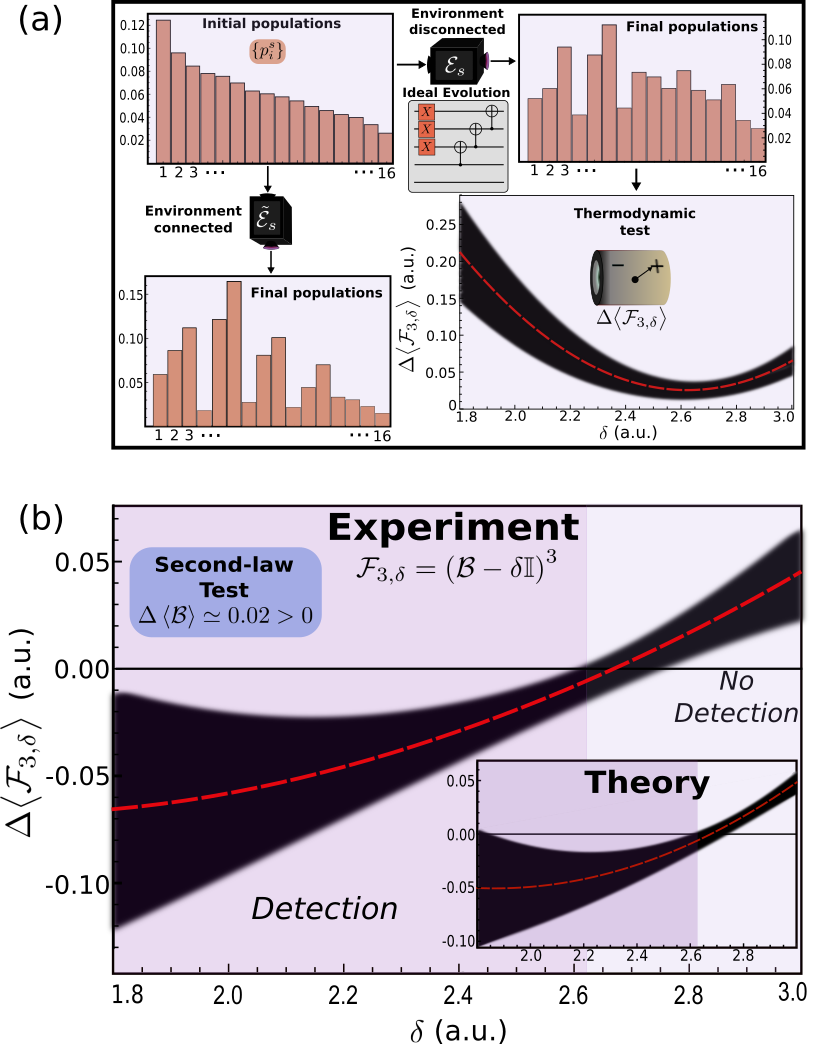}\caption{Heat leak tests performed in the Melbourne processor.\textbf{ }(a)
All the histograms depict measured system populations, averaged over
the ten batches of preparation or evolution stages (each number $1\protect\leq i\protect\leq16$
labels the same eigenstate in all histograms). The top left histogram
shows the initial populations, sorted in decreasing order. The lower
histogram shows the final populations when the environment coupling
is switched on. If the environment is decoupled the final populations
shown in the top right histogram are obtained. As expected, in this
case no heat leak is observed using the test $\Delta\left\langle \mathcal{F}_{3,\delta}\right\rangle =\Delta\left\langle (\mathcal{B}-\delta\mathbb{I})^{3}\right\rangle $
(red dashed line), within a confidence interval of three standard
deviations (shaded region). (b) When the environment is coupled, the
same test yields detection in the interval $1.8\lesssim\delta\lesssim2.6$
(same color coding of (a)). The inset shows the result for the simulation
of the ideal evolution applied to $\{p_{i}^{s}\}$. }
\end{figure}

For the experiments performed with the Melbourne processor, the preparation
and evolution stages are illustrated in Fig. 1(b). In this case we
study a heat leak acting on a four-qubit system, due to the coupling
with a single-qubit environment. The $i$th qubit in the system has
Hamiltonian $H_{i}=|1\rangle_{i}\langle1|$, where $|1\rangle_{i}$
is the excited state in the corresponding computational basis (setting
the ground energy equal to zero). The total Hamiltonian of the system
is simply $H_{s}=\sum_{i=9}^{12}H_{i}$. Accordingly, energy measurements
are associated to measurements in the four-qubit computational basis
$\{\otimes_{i=9}^{12}|j\rangle_{i}\}_{j=0,1}$. By default, all the
qubits in the processor start in the ground state. This implies\textcolor{green}{{}
}that direct preparation of mixed states is not possible. We circumvent
this limitation by employing the qubits 3-6 and 13 as ancillae, to
prepare the initial mixed state $\rho_{se}$. The procedure is indicated
in Fig 1(b). 

The IBMQE processors are subjected to gate errors and readout errors.
In the case of the Melbourne processor, the employment of cnot gates
for the preparation and the evolution introduces significant deviations
from the ideal circuits. However, we certify that the initial state
$\rho_{s}$ is well approximated by a product of thermal states with
ground populations (with $p_{0}^{(i)}$ the ground population of qubit
$i$) $p_{0}^{(12)}=0.612$, $p_{0}^{(11)}=0.586$, $p_{0}^{(10)}=0.611$,
$p_{0}^{(9)}=0.557$, and that the environment is prepared in a thermal
state $\rho_{e}$ with ground population $p_{0}^{e}=0.782$ (see Supplemental
Material \citep{suppmaterial} for further details). Since $\rho_{s}$
is compatible with Eq. (\ref{eq:1}), its deviation from the initial
state programmed in the IBMQE software interface is irrelevant for
the task of heat leak detection. To obtain the populations of $\rho_{s}$
and $\rho_{e}$, the readout error is modeled through a measurement
matrix $\mathbb{M}$ that transforms the vector of actual (without
readout error) populations $\{p_{i}^{s}\}=\{\langle i|_{s}\rho_{s}|i\rangle_{s}\}$
into observed (with readout error) populations $\mathbb{M}\{p_{i}^{s}\}$
(details about the experimental determination of this matrix are given
in \citep{suppmaterial}). In this way, the actual populations are
estimated by applying the inverse of $\mathbb{M}$ to the observed
populations. 
\begin{figure}
\centering{}\includegraphics[scale=0.65]{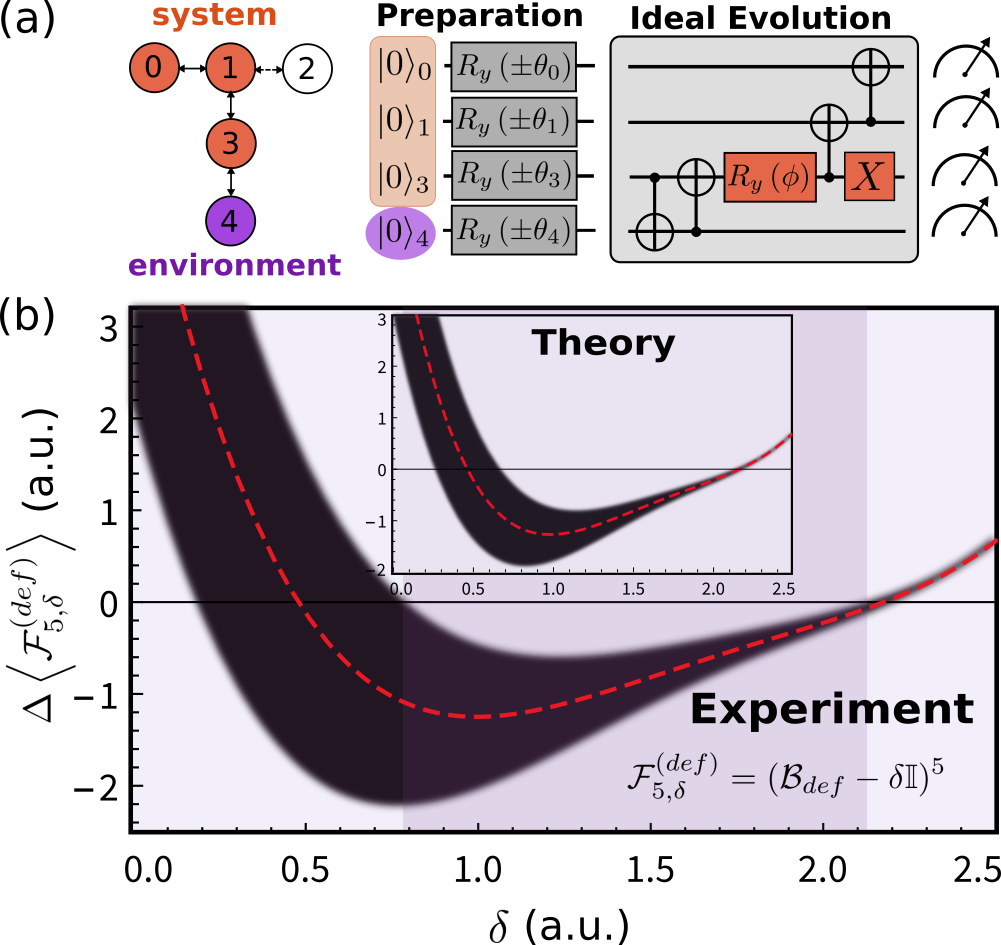}\caption{Heat leak tests performed in the five-qubit Essex processor. (a) We
employ the qubits 0, 1, and 3 as system, and the qubit 4 as environment.
The initial mixed state is prepared through single-qubit rotations
$R_{y}(\pm\theta_{i})$, using the angles $\{\theta_{0},\theta_{1},\theta_{3},\theta_{4}\}=\{0.3\pi,0.4\pi,0.4\pi,0.15\pi\}$
(see text and \citep{suppmaterial}). (b) Heat leak test based on
the observable $\mathcal{F}_{5,\delta}^{(def)}$, constructed from
a passivity deformation $\mathcal{B}_{def}$ specified in the main
text. This test detects the heat leak associated with the circuit
in (a) within three standard deviations from $\Delta\left\langle \mathcal{F}_{5,\delta}^{(def)}\right\rangle $.
The inset shows the result corresponding to the numerical simulation.}
\end{figure}

To construct the observable $\mathcal{B}=\sum_{i=1}^{16}-\textrm{ln}(p_{i}^{s})|i\rangle_{s}\langle i|$
and any passive observable ${\normalcolor {\color{red}{\normalcolor \mathcal{F}_{\alpha,\delta}}}}$
we use the initial the populations $\{p_{i}^{s}\}$. As previously
mentioned, due to the presence of gate errors the ideal evolution
$U_{se}^{\textrm{id}}$ coded in the software interface is effectively
implemented as a map ${\color{red}{\normalcolor \tilde{\mathcal{E}}_{se}}}$
that yields the experimental final state ${\normalcolor {\color{red}{\normalcolor \rho'_{s}=\textrm{Tr}_{e}[\tilde{\mathcal{E}}_{se}(\rho_{se})]\equiv\tilde{\mathcal{E}}_{s}(\rho_{s})}}}$.
The corresponding final distribution $\{p'{}_{i}^{s}\}$ is also estimated
by applying the inverse of the measurement matrix to the observed
distribution. In this way, the test $\textrm{Tr}[{\color{red}{\normalcolor \mathcal{F}_{\alpha,\delta}}}(\rho'_{s}-\rho_{s})]$
is evaluated as 
\begin{equation}
\Delta\left\langle \mathcal{F}_{\alpha,\delta}\right\rangle =\sum_{i=1}^{16}f_{\alpha,\delta}^{(i)}(p'{}_{i}^{s}-p{}_{i}^{s}),\label{eq:4}
\end{equation}
where $f_{\alpha,\delta}^{(i)}=(\mathcal{B}_{i}-\delta)^{\alpha}$
and $\mathcal{B}_{i}=-\textrm{ln}(p_{i}^{s})$. The experimental data
are collected by implementing the circuits that include the preparation
and evolution stages. We implement ten batches for each circuit, with
each batch being composed of 8192 shots. 

Figure 2 presents the experimental results of a heat leak test based
on the observable $\mathcal{F}_{3,\delta}$. The histograms in Fig.
2(a) depict the initial populations $\{p_{i}^{s}\}$ and final populations
$\{p'{}_{i}^{s}\}$, given by normalized count frequencies from a
total sample of $10\times8192$ shots. The red dashed curves in Fig.
2 are obtained by plugging these populations into Eq. (\ref{eq:4}).
Moreover, shaded regions in each plot stand for a confidence interval
of three standard deviations (see \citep{suppmaterial}). In Fig.
2(a) it is shown that the test $\Delta\left\langle \mathcal{F}_{3,\delta}\right\rangle $
yields only positive values when the environment is decoupled, as
expected. When the environment is coupled, Fig. 2(b) shows detection
of the heat leak with the same test, for $1.8\lesssim\delta\lesssim2.6$.
Importantly, the application of this test is motivated by the fact
that unambiguous detection (within the confidence interval) is not
possible with the observable $\mathcal{F}_{2,\delta}$, based on the
smaller power $\alpha=2$. The corresponding plot is provided in \citep{suppmaterial}.
Moreover, for $\alpha=1$ the test $\Delta\left\langle \mathcal{F}_{1,\delta}\right\rangle =\Delta\left\langle \mathcal{B}\right\rangle $
also fails (see Fig. 2(b)). This clearly demonstrates a situation
where global passivity outperforms the standard second law (\ref{eq:2}),
regarding detection sensitivity. In \citep{suppmaterial} it is also
shown that for $0\leq\alpha\leq4$ no detection occurs with the observables
$\mathcal{B}^{\alpha}$. 

Now we consider an experiment where passivity deformation outperforms
both the second law and global passivity. Passivity deformation is
a method to systematically construct passive observables, by performing
suitable transformations on the eigenvalues of $\mathcal{B}$. In
this work we employ the special deformation 
\begin{equation}
\mathcal{B}\rightarrow\mathcal{B}_{def}=\sum_{i}\beta_{i}^{(def)}H_{i},\label{eq:5}
\end{equation}
where $\beta_{i}^{(def)}$ are effective inverse temperatures that
guarantee the passivity of $\mathcal{B}_{def}$. Figure 3(a) illustrates
the structure of the Essex processor and the circuit used for the
implementation of the heat leak, which involves a three-qubit system
that interacts with a single-qubit environment. As explained below,
detection with the test shown in Fig. 3(b) is possible thanks to the
employment of a suitable deformation. 

The limited size of the Essex processor prevents us to prepare the
initial state $\rho_{s}\otimes\rho_{e}$ through entanglement with
ancillae. However, a diagonal state of the form (\ref{eq:1}) can
be obtained from an ensemble of coherent states with identical populations
in the energy basis. We first separately prepare sixteen coherent
states, by applying single-qubit rotations $\{R_{y}(\pm\theta_{i})\}_{i=0,1,3,4}$,
see Fig. 3(a). As explained in \citep{suppmaterial}, by mixing these states
with equal probabilities we obtain a product of thermal states with
ground populations $p_{0}^{(i)}=\textrm{cos}^{2}(\theta_{i}/2)$.
In this case four batches of 8192 shots are employed for each initial
coherent state and its final counterpart. 

In addition to the test shown in Fig. 3(b), we confirmed that the
heat leak is not detected with the observable $\mathcal{B}$, neither
with tests based on the observables $\mathcal{F}_{\alpha,\delta}=(\mathcal{B}-\delta\mathbb{I})^{\alpha}$,
for $\alpha=2,3,4,5$. This difficulty is overcome through a deformation
of $\mathcal{B}$, characterized by the transformation $\{\beta_{0},\beta_{1},\beta_{3}\}\rightarrow\{\beta_{0},0,\beta_{0}\}$.
The resulting observable $\mathcal{B}_{def}=\beta_{0}(H_{1}+H_{3})$
is equivalent to the total Hamiltonian of the qubits 1 and 3. As occurs
with the non-deformed observables $\mathcal{F}_{\alpha,\delta}$,
the deformation $\mathcal{B}_{def}$ gives rise to a family of observables
$\{\mathcal{F}_{\alpha,\delta}^{(def)}\}\equiv\{(\mathcal{B}_{def}-\delta\mathbb{I})^{\alpha}\}_{\alpha\geq0}$
which are passive if the shift $\delta$ is properly chosen. These
observables also exhibit a polynomial scaling similar to that described
for $\mathcal{F}_{\alpha,\delta}$ (see \citep{suppmaterial}), with $n$
replaced by an effective number of qubits $n_{def}$ that satisfies
$n_{def}\leq n$. 

In Fig. 3(b) we use the same color coding as in Fig. 2(c). Clearly,
for shift values $0.8\lesssim\delta\lesssim2.1$ detection takes place
with the deformed observable $\mathcal{F}_{5,\delta}^{(def)}$. In
\citep{suppmaterial} we also show that if the environment is decoupled
the test $\Delta\left\langle \mathcal{F}_{5,\delta}^{(def)}\right\rangle $
is positive, as expected. This provides additional evidence that the
source of the heat leak is the engineered environment interaction
and not some intrinsic imperfection in the processor. 

Finally, we remark that the advantage of shifted observables and deformed
observables can be understood by explicitly writing the expansion
\begin{equation}
\Delta\left\langle (\mathcal{B}-\delta\mathbb{I})^{\alpha}\right\rangle =\sum_{k=0}^{\alpha}\left(\begin{array}{c}
\alpha\\
k
\end{array}\right)(-1)^{k}\delta^{k}\Delta\left\langle \mathcal{B}^{\alpha-k}\right\rangle ,\label{eq:6}
\end{equation}
which is valid for $\alpha$ positive and integer. Since for $k$
odd the factor $\left(\begin{array}{c}
\alpha\\
k
\end{array}\right)(-1)^{k}\delta^{k}$ is negative, the test $\Delta\left\langle (\mathcal{B}-\delta\mathbb{I})^{\alpha}\right\rangle $
can yield detection even if all the non-shifted tests $\Delta\left\langle \mathcal{B}^{\alpha-k}\right\rangle $
are positive. However, detection also requires that more weight is
given to the negative factors than to the positive ones (associated
with $\alpha$ even). The deformation $\mathcal{B}\rightarrow\mathcal{B}_{def}$
provides the appropriate weights $\Delta\left\langle \mathcal{B}_{def}^{\alpha-k}\right\rangle $
for this to happen, in the case of the second circuit analyzed. 

\textit{Discussion}. The TUR used in Ref. \citep{buffoni2020thermodynamics}
allows to conclude that for a certain annealing protocol the device
operates as a thermal accelerator, if the annealer is prepared in
a hot thermal state. Therefore, it can be asserted that the annealer
does not evolve unitarily. This result relies on preliminary evidence
that the background cold bath indeed affects the system dynamics.
In particular, the application of the TUR requires to estimate first
the temperature of this bath. Although we perform thermodynamic tests
using an engineered enviroment, we stress that they provide unambiguous
detection without any prior information on the potential source of
noise. 

Resource theory (RT) also provides an infinite set of thermodynamic
constraints for systems coupled to a thermal bath through energy-preserving
interactions \citep{Brandao2015}. In \citep{suppmaterial} we show an
example where no constraint from this infinite familiy can be violated
by the presence of a hidden environment. Specifically, we study a
four-qubit system that is divided into a ``bath'', given by a subsystem
at fixed temperature, and the remaining subsystem where the constraint
is examined. The total system evolves through an energy-preserving
circuit and an external interaction with a qubit environment. Importantly,
even in cases where violations of RT constraints are observed they
can be due to external classical drivings, and therefore such violations
can also occur for unitary dynamics. For the studied example, we also
show that the environment is detected using the passive observable
$\mathcal{F}_{7,\delta}$. 

\textit{Conclusions}. In this work we experimentally show that thermodynamic
inequalities recently derived \citep{uzdin2018global,uzdin2019passivity}
are useful for the detection of heat leaks in quantum circuits. While
the experiments are performed on small size circuits, we theoretically
demonstrate efficient scalability to larger devices. In this context,
error detection and characterization faces substantial experimental
and computational challenges. These challenges are ultimately rooted
in the exponential growth of resources needed to accurately estimate
a quantum state or to simulate its evolution \citep{arute2019quantum,boixo2018characterizing,zhong2020quantum,haah2017sample}.
However, several recent results show that certain quantum properties
can be efficiently estimated, i.e. without incurring in an exponential
overhead \citep{huang2020predicting,aaronson2019shadow,paini2020estimating}.
The efficiency of the thermodynamic tests studied here builds up on
such a possibility. Although they are specifically designed to detect
non-unital dynamics, they are economic not only in terms of measurements,
but also by saving the computational power required to compare experimental
and simulated evolutions. A current limitation of these tests is that
they cannot detect any non-unital error, see \citep{suppmaterial}. In
addition, finding proper passive observables for detection in specfic
situations is a non-trivial problem. These sensitity constraints can
be interpreted as a ``price to pay'' in exchange for scalability,
and further research concerning this trade-off is an interesting open
problem. Apart from heat leaks, a pertinent question is if thermodynamic
tools can be applied for diagnostics of other relevant errors, such
as dephasing. 

\section*{Supplemental Material}

\global\long\def\thesection{S-\Roman{section}}
\setcounter{section}{0}
\global\long\def\thefigure{S\arabic{figure}}
\setcounter{figure}{0}
\global\long\def\theequation{S\arabic{equation}}
\setcounter{equation}{0}

\section{Scaling of different heat leak tests with respect to system size }

In the following we show that heat leak tests based on passive observables
constitute an efficient method for the detection of non-unital errors
in quantum devices. While in theory these errors can also be diagnosed
using other more direct methods, we will also explain why such strategies
are inefficient and become infeasible as the size of the system increases.
More precisely, 
\begin{itemize}
\item Direct methods such as quantum state tomography or quantum process
tomography involve a number of measurements that grow exponentially
with respect to the number of qubits $n$ in the system \citep{haah2017sample,o2016efficient}.
Such an exponential scaling is what we refer to as ``inefficient''. 
\item Conversely, heat leak tests using passive observables are ``efficient''
in the sense that their measurement cost is polynomial with respect
to $n$. Accordingly, these tests can be a more appealing and practical
alternative for diagnostics of non-unital errors in large devices.
\end{itemize}
The basic tool to demonstrate the aforementioned efficiency is a theoretic-information
result recently derived \citep{huang2020predicting}. This result
refers to the number of measurements $N$ required to accurately estimate
the mean values of a given set of $M$ observables $\{O_{i}\}_{1\leq i\leq M}$.
Given a general quantum state state $\rho$, 
\begin{equation}
N\sim\mathcal{O}\left(\frac{\textrm{log}(M/\mu)}{\epsilon^{2}}\textrm{max}_{i}\left\Vert O_{i}\right\Vert _{\textrm{shadow}}^{2}\right),\label{eq:20 N}
\end{equation}
measurements suffice to estimate \textit{all} the mean values $\bigl\langle O_{i}\bigr\rangle=\textrm{Tr}[\rho O_{i}]$
with maximum error $\epsilon$. Specifically, in \citep{huang2020predicting}
it is shown that, if $N$ satisfies Eq. (\ref{eq:20 N}), there is
an explicit classical protocol that yields estimations $o_{i}$ of
$\bigl\langle O_{i}\bigr\rangle$, such that $|o_{i}-\bigl\langle O_{i}\bigr\rangle|\leq\epsilon$
for $1\leq i\leq M$, with success probability $\textrm{Pr}\left(|o_{i}-\bigl\langle O_{i}\bigr\rangle|\leq\epsilon\right)\geq1-\mu$.
Equivalently, the protocol achieves $|o_{i}-\bigl\langle O_{i}\bigr\rangle|\leq\epsilon$
with maximum failure probability $\mu$.

The observables $O_{i}$ can be arbitrary and the ``shadow norm''
$\left\Vert \cdot\right\Vert _{\textrm{shadow}}$ is determined by
the specific measurement procedure. For passive observables derived
from GP and PD, we show in Section B of this appendix that the corresponding
estimation can be obtained from the simultaneous estimation of $M=n\frac{n^{\alpha}-1}{n-1}$
\textit{local} observables $O_{i}$, where $\alpha$ is the maximum
number of qubits on which these observables act non-trivially. In
this case, a suitable measurement procedure (explained in Section
B) yields a norm $\left\Vert O_{i}\right\Vert _{\textrm{shadow}}^{2}$
whose maximum depends only on $\alpha$. Thus, we can combine Eq.
(\ref{eq:20 N}) with error propagation, to show in Section B that
estimating the mean value of passive observables involves the polynomial
measurement cost given in Eq. (\ref{eq:22 N'}). This implies that
the corresponding heat leak tests can also be efficiently evaluated. 

\subsection{Direct methods for the detection of heat leaks (non-unital errors)
and experimental cost}

\textbf{Detection via quantum state tomography}. As explained in the
main text, heat leaks are associated with transformations $\rho\rightarrow\rho'$
that cannot be written as $\rho'=\sum_{k}\text{\ensuremath{\lambda}}_{k}U_{k}\rho U_{k}^{\dagger}$,
where $\{\lambda_{k}\}$ are probabilities and $\{U_{k}\}$ are unitary
maps. The criterion of majorization provides a theoretical tool to
directly check if $\rho\rightarrow\rho'$ contains a heat leak. Specifically,
let $\{r_{i}^{\downarrow}\}$ and $\{r'{}_{i}^{\downarrow}\}$ denote
respectively the eigenvalues of $\rho$ and $\rho'$, arranged in
non-increasing order: $r_{i}^{\downarrow}\geq r_{i+1}^{\downarrow}$
and $r_{i}'{}^{\downarrow}\geq r'_{i+1}{}^{\downarrow}$ for all $1\leq i\leq d$.
The state $\rho$ majorizes the state $\rho'$, denoted as $\rho\succ\rho'$,
iff $S_{j}\equiv\sum_{i=1}^{j}r_{i}^{\downarrow}\geq S'_{j}\equiv\sum_{i=1}^{j}r'{}_{i}^{\downarrow}$
for all $1\leq j\leq d$ \citep{nielsen2002introduction,marshall1979inequalities}.
Since $\rho\succ\rho'$ iff there exist $\{\lambda_{k}\}$ and $\{U_{k}\}$
such that $\rho'=\sum_{k}\text{\ensuremath{\lambda}}_{k}U_{k}\rho U_{k}^{\dagger}$
\citep{nielsen2002introduction}, \textit{any} heat leak manifests
itself in an inequality $S_{j}<S'_{j}$, for some $1\leq j\leq d$.
However, knowing the partial sums $\{S'_{j}\}_{1\leq j\leq d}$ requires
to know both the initial and final eigenvalues $\{r_{i}{}^{\downarrow}\}$
and $\{r_{i}'{}^{\downarrow}\}$. 

The experimental determination of $\{r'_{i}{}^{\downarrow}\}$ involves
in the worst case quantum state tomography (state tomography for simplicity)
of the final state $\rho'$, which consists in a full experimental
reconstruction of $\rho'$. Of course this is also true for the eigenvalues
of the initial state $\rho$. From fundamental information-theoretic
bounds it is known that at least $N\sim\mathcal{O}(\textrm{rank}(\rho')d)$
measurements are required for state tomography \citep{haah2017sample}.
For example, if $\rho'$ describes a system of $n$ qubits, the number
of measurements needed grows exponentially with $n$. In particular,
this makes state tomography infeasible for detecting heat leaks in
quantum devices sufficiently large to perform useful tasks. 

\textbf{Detection via population estimation in the computational basis}.
Some heat leaks could in principle be diagnosed by checking a violation
of majorization between the initial state and the diagonal part of
$\rho'$ in a suitable basis. In a quantum processor, the natural
choice to measure the diagonal of $\rho'$ is the computational basis,
where measurements can be directly performed (measurements in other
bases involve pre-measurement gates that can introduce additional
errors). Let $\{|i_{\textrm{c}}\rangle\}_{1\leq i\leq d}$ denote
the computational basis, and $s_{i}=\langle i_{\textrm{c}}|\rho'|i_{\textrm{c}}\rangle$
the $i$th diagonal element of $\rho'$ in this basis. Moreover, let
$D(\rho')$ be the diagonal matrix whose entries are given by the
$s_{i}$. Since $D(\rho')$ is simply $\rho'$ dephased in the computational
basis, it can be written as a mixture of unitaries acting on $\rho'$
\citep{watrous2018theory}, which also implies that $\rho'\succ D(\rho')$.
From transitivity of majorization, it follows that if $\rho\succ\rho'$
then $\rho\succ D(\rho')$.\textcolor{red}{{} }

Using the populations $s_{i}$ one can detect a heat leak if a relation
of the form $\sum_{i=1}^{j}r_{i}^{\downarrow}<\sum_{i=1}^{j}s_{i}^{\downarrow}$
holds, for some $1\leq j\leq d$. While this technique is clearly
more economic than state tomography of $\rho'$, for a system of $n$
qubits a reconstruction of all the partial sums $\sum_{i=1}^{j}s_{i}^{\downarrow}$
still involves measuring $2^{n}-1$ populations $s_{i}$ (with one
of them deduced by normalization). In the following appendix we will
discuss in more detail the experimental cost of this method, since
there is a connection between the inequalities $\sum_{i=1}^{j}r_{i}^{\downarrow}<\sum_{i=1}^{j}s_{i}^{\downarrow}$
and the heat leaks that can be detected using passive observables.
In particular, we will show that even for a single value of $j$ it
is necessary to estimate all the final populations $s_{i}$, to unambiguosly
evaluate the partial sum $\sum_{i=1}^{j}s_{i}^{\downarrow}$. 

\textbf{Quantum process tomography and transfer matrix}. Perhaps the
most direct route for error characterization in any quantum device
would be to perform quantum process tomography, or process tomography
for brevity. This consists in the full experimental characterization
of a quantum process. If this process is represented by a quantum
channel $\mathcal{E}$ that maps density matrices into density matrices,
process tomography aims at determining the action of $\mathcal{E}$
on any possible initial state $\rho$. To that end the channel must
be implemented on $d^{2}$ states that form a tomographically complete
basis \citep{nielsen2002quantum}, and state tomography must be applied
on each of the corresponding $d^{2}$ outputs. From the previous discussion
on state tomography it is clear that the scaling of process tomography
is even worse. The discussion regarding the inefficiency of population
measurements leads to an analogous conclusion for the transfer matrix,
which may be seen a coarse-grained version of $\mathcal{E}$. The
transfer $T$ matrix is $d\times d$ matrix with elements $T_{ij}\equiv\textrm{Tr}\left[|i_{\textrm{c}}\rangle\langle i_{\textrm{c}}|\mathcal{E}(|j_{\textrm{c}}\rangle\langle j_{\textrm{c}}|)\right]$.
To reconstruct $T$, the process $\mathcal{E}$ must be implemented
on the $d=2^{n}$ computational eigenstates and the final populations
$T_{ij}$ have to be measured. This implies that measuring $T$ is
not a practical alternative for error detection even for systems of
moderate size. 

\subsection{Efficient evaluation of heat leak tests based on passive observables}

To understand why passive observables $\mathcal{F}=\mathcal{F}_{\alpha,\delta}$
and $\mathcal{F}=\mathcal{F}_{\alpha,\delta}^{(def)}$ are an efficient
tool for heat leak detection we must first describe the relevant parameters
entering the operator norm $\left\Vert \cdot\right\Vert _{\textrm{shadow}}$
in Eq. (\ref{eq:20 N}). As a matter of fact, we will see that for
passive observables this norm is upper bounded by a constant that
is independent of $d$. In what follows we refer to the final state
$\rho'$, which yields the final mean value $\bigl\langle\mathcal{F}\bigr\rangle'=\textrm{Tr}(\mathcal{F}\rho')$.
However, the same arguments are valid for the initial mean value $\bigl\langle\mathcal{F}\bigr\rangle=\textrm{Tr}(\mathcal{F}\rho)$. 

The experimental procedure underlying Eq. (\ref{eq:20 N}) is based
on measuring $\rho'$ in $N$ randomly chosen measurement bases \citep{huang2020predicting}.
These bases are determined by randomly selecting unitaries $U$ from
some ensemble $\mathcal{U},$and then measuring the transformed state
$U\rho'U^{\dagger}$ in the computational basis (this effectively
implements a measurement in the rotated basis $\{U^{\dagger}|i_{\textrm{c}}\rangle\}$).
Afterwards, a classical postprocessing is applied to the $N$ outcomes
to obtain estimated mean values $\{o_{i}\}_{1\leq i\leq M}$ with
maximum error $\epsilon$ and maximum failure probability $\mu$. 

\textbf{Lemma 1 (based on the results from }\citep{huang2020predicting}\textbf{)}.
The norm $\left\Vert \cdot\right\Vert _{\textrm{shadow}}$ depends
on the ensemble of unitaries $\mathcal{U}$. In Ref. \citep{huang2020predicting}
the authors consider two ensembles: arbitrary Clifford unitaries acting
on $n$ qubits, and tensor products of single-qubit Clifford unitaries.
In the second case the ($n$-qubit) measurement bases are random tensor
products of Pauli bases $\{X,Y,Z\}$. For this choice, \textit{the
norm $\left\Vert \cdot\right\Vert _{\textrm{shadow}}$ depends only
on the locality of the observable $O$ and not on the dimension $d$}.
Specifically, for an observable $O=\tilde{O}\otimes\mathbb{I}^{n-k}$,
where $\tilde{O}$ acts non trivially on $k$ qubits and $\mathbb{I}^{n-k}$
is the identity on the remaining $n-k$ qubits, $\left\Vert O\right\Vert _{\textrm{shadow}}^{2}\leq4^{k}\left\Vert O\right\Vert _{\infty}^{2}$,
being $\left\Vert \cdot\right\Vert _{\infty}$ the spectral norm (Proposition
3 in \citep{huang2020predicting}). 

A passive observable $\mathcal{F}_{\alpha,\delta}$ is a sum of local
observables acting on $k\leq\alpha$ qubits. This is deduced by explicitly
writing the binomial expansion for $\mathcal{F}_{\alpha,\delta}=(\mathcal{B}-\delta\mathbb{I})^{\alpha}$:
\begin{align}
\mathcal{F}_{\alpha,\delta} & =(\mathcal{B}-\delta\mathbb{I})^{\alpha}\nonumber \\
 & =\sum_{k=0}^{\alpha}\left(\begin{array}{c}
\alpha\\
k
\end{array}\right)(-1)^{\alpha-k}\delta^{\alpha-k}\mathcal{B}^{k}.\label{eq:20.1 passive observable}
\end{align}
From the definition of $\mathcal{B}$, $\mathcal{B}=\sum_{i=1}^{n}\beta_{i}H_{i}$,
we also have that 
\begin{align*}
\mathcal{B}^{k} & =\left(\sum_{i=1}^{n}\beta_{i}H_{i}\right)^{k}\\
 & =\sum_{i_{1}=1}^{n}\sum_{i_{2}=1}^{n}...\sum_{i_{k}=1}^{n}\left(\otimes_{m=1}^{k}\beta_{i_{m}}H_{i_{m}}\right)\\
 & =\sum_{\{i_{m}\}}\left(\prod_{m=1}^{k}\beta_{i_{m}}\right)\tilde{\mathcal{H}}_{k,\{i_{m}\}},
\end{align*}
where $\tilde{\mathcal{H}}_{k,\{i_{m}\}}\equiv\otimes_{m=1}^{k}H_{i_{m}}$and
the sum $\sum_{\{i_{m}\}}$ is over the $n^{k}$ sets of indices $\{i_{m}\}_{1\leq m\leq k}$
that label groups of $k$ qubits. Therefore, $\mathcal{B}^{k}$ is
a sum of $n^{k}$ observables that act locally on sets of $k$ qubits. 

An estimation of the mean value $\bigl\langle\mathcal{F}_{\alpha,\delta}\bigr\rangle$
can be constructed from the estimation of all the mean values $\bigl\langle\tilde{\mathcal{H}}_{k,\{i_{m}\}}\bigr\rangle$.
This has to be done carefully, keeping in mind that the errors for
the $\bigl\langle\tilde{\mathcal{H}}_{k,\{i_{m}\}}\bigr\rangle$ propagate
to $\bigl\langle\mathcal{F}_{\alpha,\delta}\bigr\rangle$. However,
we show that if $N$ measurements allow to estimate $\left\{ \bigl\langle\tilde{\mathcal{H}}_{k,\{i_{m}\}}\bigr\rangle\right\} $
with accuracy $(\epsilon,\mu)$, the same accuracy can be achieved
on $\bigl\langle\mathcal{F}_{\alpha,\delta}\bigr\rangle$ if $N$
is increased to $N'$, with an increment that is only polynomial in
$n$. Accordingly, efficient estimation of $\left\{ \bigl\langle\tilde{\mathcal{H}}_{k,\{i_{m}\}}\bigr\rangle\right\} $
implies efficient estimation of $\bigl\langle\mathcal{F}_{\alpha,\delta}\bigr\rangle$.
We show first that 
\begin{equation}
N\sim\mathcal{O}\left(\frac{\textrm{log}(n/\mu)}{\epsilon^{2}}\right).\label{eq:21 N for H_k,=00007Bim=00007D}
\end{equation}
To this end we use the fact that the norm $\left\Vert \tilde{\mathcal{H}}_{k,\{i_{m}\}}\right\Vert _{\textrm{shadow}}^{2}$
satisfies $\left\Vert \tilde{\mathcal{H}}_{k,\{i_{m}\}}\right\Vert _{\textrm{shadow}}^{2}\leq4^{k}$,
which follows from Lemma 1 and the fact that $\left\Vert \otimes_{m=1}^{k}H_{i_{m}}\right\Vert _{\infty}=1$
(since by convention the maximum eigenvalue of each $H_{i_{m}}$ is
1). Hence, $\textrm{max}_{1\leq k\leq\alpha,\{i_{m}\}}\left\Vert \tilde{\mathcal{H}}_{k,\{i_{m}\}}\right\Vert _{\textrm{shadow}}^{2}$
is independent of $n$ and can be absorbed in the implicit constants
of Eq. (\ref{eq:20 N}). On the other hand, there are $n^{k}$ observables$\tilde{\mathcal{H}}_{k,\{i_{m}\}}$
for $k$ fixed, which yields the total number 
\[
M=\sum_{k=1}^{\alpha}n^{k}=n^{\alpha}\sum_{k=0}^{\alpha-1}\left(\frac{1}{n}\right)^{k}=n\frac{n^{\alpha}-1}{n-1}.
\]
For $n$ large, it is straightforward to check that $\textrm{log}(M)\sim\alpha\textrm{log}(n)$.
By inserting this into Eq. (\ref{eq:20 N}) we obtain Eq. (\ref{eq:21 N for H_k,=00007Bim=00007D}). 

Now we proceed to determine the (order of the) number of measurements
$N'$ to estimate $\bigl\langle\mathcal{F}_{\alpha,\delta}\bigr\rangle$
keeping accuracy $(\epsilon,\mu)$. Let $\tilde{H}_{k,\{i_{m}\}}$
and $F_{\alpha,\delta}$ denote respectively the estimations of the
mean values $\bigl\langle\tilde{\mathcal{H}}_{k,\{i_{m}\}}\bigr\rangle$
and $\bigl\langle\mathcal{F}_{\alpha,\delta}\bigr\rangle$. If we
require that $\bigl|\tilde{H}_{k,\{i_{m}\}}-\bigl\langle\tilde{\mathcal{H}}_{k,\{i_{m}\}}\bigr\rangle\bigr|\leq\epsilon$
with probability $\textrm{Pr}\left(\bigl|\tilde{H}_{k,\{i_{m}\}}-\bigl\langle\tilde{\mathcal{H}}_{k,\{i_{m}\}}\bigr\rangle\bigr|\leq\epsilon\right)\geq1-\mu$
for all $k,\{i_{m}\}$, then $\bigl|F_{\alpha,\delta}-\bigl\langle\mathcal{F}_{\alpha,\delta}\bigr\rangle\bigr|\leq\epsilon_{\textrm{tot}}$
with probability $\textrm{Pr}\left(|F_{\alpha,\delta}-\bigl\langle\mathcal{F}_{\alpha,\delta}\bigr\rangle|\leq\epsilon_{\textrm{tot}}\right)\geq1-\mu$.
The total error $\epsilon_{\textrm{tot}}$ is additive because $\mathcal{F}_{\alpha,\delta}$
is a linear combination of $\tilde{\mathcal{H}}_{k,\{i_{m}\}}$ and
this linearity extends to the mean value. That is, 
\begin{equation}
\bigl\langle\mathcal{F}_{\alpha,\delta}\bigr\rangle=\sum_{k=0}^{\alpha}\sum_{\{i_{m}\}}c_{k,\{i_{m}\}}^{(\alpha,\delta)}\bigl\langle\tilde{\mathcal{H}}_{k,\{i_{m}\}}\bigr\rangle,\label{eq:21.05 mean value of F in terms of <H_k,=00007Bim=00007D>}
\end{equation}
where $c_{k,\{i_{m}\}}^{(\alpha,\delta)}\equiv\left(\begin{array}{c}
\alpha\\
k
\end{array}\right)(-\delta)^{\alpha-k}\prod_{m=1}^{k}\beta_{i_{m}}$. Next, we multiply each inequality $\bigl|\tilde{H}_{k,\{i_{m}\}}-\bigl\langle\tilde{\mathcal{H}}_{k,\{i_{m}\}}\bigr\rangle\bigr|\leq\epsilon$
by the coefficient $c_{k,\{i_{m}\}}^{(\alpha,\delta)}$ and take the
sum over $k$ and $\{i_{m}\}$, which leads to the expression $\bigl|F_{\alpha,\delta}-\bigl\langle\mathcal{F}_{\alpha,\delta}\bigr\rangle\bigr|\leq\sum_{k,\{i_{m}\}}\bigl|c_{k,\{i_{m}\}}^{(\alpha,\delta)}\bigr|\epsilon$.
The quantity $\sum_{k,\{i_{m}\}}\bigl|c_{k,\{i_{m}\}}^{(\alpha,\delta)}\bigr|\epsilon$
thus determines the maximum total error $\epsilon_{\textrm{tot}}$.
By applying the bound $c_{k,\{i_{m}\}}^{(\alpha,\delta)}\leq\textrm{max}_{k,\{i_{m}\}}\left(\prod_{m=1}^{k}\beta_{i_{m}}\right)\left(\begin{array}{c}
\alpha\\
k
\end{array}\right)(-\delta)^{\alpha-k}$, we conclude that 
\begin{align}
\frac{\epsilon_{\textrm{tot}}}{\epsilon} & \leq\textrm{max}_{\{i_{m}\}}\left(\prod_{m=1}^{k}\beta_{i_{m}}\right)\sum_{k=0}^{\alpha}\left(\begin{array}{c}
\alpha\\
k
\end{array}\right)(\delta)^{\alpha-k}n^{k}\nonumber \\
 & =\textrm{max}_{\{i_{m}\}}\left(\prod_{m=1}^{k}\beta_{i_{m}}\right)(n+\delta)^{\alpha}.\label{eq:21.1 total error}
\end{align}

Equation (\ref{eq:21.1 total error}) indicates that if each estimated
value $\tilde{H}_{k,\{i_{m}\}}$ has maximum error $\epsilon$, the
maximum error corresponding to $F_{\alpha,\delta}$ has an upper bound
that grows polynomially in $n$. On the other hand, the minimum success
probability is reduced to $1-\mu_{\textrm{tot}}=(1-\mu)^{n\frac{n^{\alpha}-1}{n-1}}$.
This decrement in accuracy for $F_{\alpha,\delta}$ can be compensated
by imposing a higher accuracy $(\epsilon',\mu')$ for the estimation
of each $\bigl\langle\tilde{\mathcal{H}}_{k,\{i_{m}\}}\bigr\rangle$,
such that $(\epsilon_{\textrm{tot}},\mu_{\textrm{tot}})=(\epsilon,\mu)$.
If we perform the substitutions $\epsilon_{\textrm{tot}}\rightarrow\epsilon$
and $\epsilon\rightarrow\epsilon'$, Eq. (\ref{eq:21.1 total error})
yields the lower bound $\epsilon'\geq\frac{\epsilon}{\textrm{max}_{\{i_{m}\}}\left(\prod_{m=1}^{k}\beta_{i_{m}}\right)(n+\delta)^{\alpha}}$.
Moreover, from $1-\mu=(1-\mu')^{n\frac{n^{\alpha}-1}{n-1}}$ we obtain
$\mu'=1-(1-\mu)^{\frac{n-1}{n(n^{\alpha}-1)}}$. The Taylor expansion
of this expression around zero yields $\mu'\sim\frac{n-1}{n(n^{\alpha}-1)}\mu$,
for $\mu\ll1$. In this way, the number of measurements required to
estimate $\bigl\langle\mathcal{F}_{\alpha,\delta}\bigr\rangle$ with
accuracy $(\epsilon,\mu)$ is 
\begin{equation}
N'\sim\mathcal{O}\left(\frac{\textrm{log}(n/\mu')}{\epsilon'^{2}}\right)=\mathcal{O}\left(\frac{\textrm{log}(n/\mu)}{\epsilon{}^{2}}(n+\delta)^{2\alpha}\right),\label{eq:22 N'}
\end{equation}
where we have made the approximation $\textrm{log}\left(\frac{n(n^{\alpha}-1)}{n-1}\right)\sim\alpha\textrm{log}(n)$.
Note also that the quantity $\textrm{max}_{\{i_{m}\}}\left(\prod_{m=1}^{k}\beta_{i_{m}}\right)$
can be absorbed with the other implicit constants that do not depend
on $n$. 

Equation (\ref{eq:22 N'}) is the main result of this appendix. It
tells us that we can accurately estimate $\bigl\langle\mathcal{F}_{\alpha,\delta}\bigr\rangle$
with a number of measurements that scales polynomially in the number
of qubits, \textit{irrespective of the state on which $\bigl\langle\mathcal{F}_{\alpha,\delta}\bigr\rangle$
is evaluated}. In addition to $F_{\alpha,\delta}$, let $F'_{\alpha,\delta}$
denote the estimation of the final mean values $\bigl\langle\mathcal{F}_{\alpha,\delta}\bigr\rangle'=\textrm{Tr}(\rho'\mathcal{F}_{\alpha,\delta})$.
If we require that $F'_{\alpha,\delta}$ also has accuracy $(\epsilon,\mu)$,
the estimation of the test $\Delta\bigl\langle\mathcal{F}_{\alpha,\delta}\bigr\rangle=\bigl\langle\mathcal{F}_{\alpha,\delta}\bigr\rangle'-\bigl\langle\mathcal{F}_{\alpha,\delta}\bigr\rangle$,
denoted as $\Delta F_{\alpha,\delta}$, has accuracy $(2\epsilon,(1-\mu)^{2})$.
Therefore, this test can also be accurately evaluated with $N'$ given
by Eq. (\ref{eq:22 N'}). 

Summarizing, the estimation of $\Delta\bigl\langle\mathcal{F}_{\alpha,\delta}\bigr\rangle$
involves two stages. A first one that could be referred to as ``direct
estimation'', which is applied to the observables $\{\tilde{\mathcal{H}}_{k,\{i_{m}\}}\}_{1\leq k\leq\alpha,\{i_{m}\}}$,
with $\tilde{\mathcal{H}}_{k,\{i_{m}\}}=\otimes_{m=1}^{k}H_{i_{m}}$
an arbitrary product of $k$ single-qubit Hamiltonians. Equation (\ref{eq:21 N for H_k,=00007Bim=00007D})
provides the order of the number of measurements required to simultaneously
produce \textit{all} the (initial and final) estimated values $\{\tilde{H}_{k,\{i_{m}\}}\}$,
with accuracy $(\epsilon,\mu)$. The second ``indirect estimation''
stage consists of using Eq. (\ref{eq:21.05 mean value of F in terms of <H_k,=00007Bim=00007D>})
to obtain $\Delta F_{\alpha,\delta}$ from $\{\tilde{H}_{k,\{i_{m}\}}\}$.
Due to error propagation, to achieve accuracy $(2\epsilon,(1-\mu)^{2})$
in the estimation of $\Delta\bigl\langle\mathcal{F}_{\alpha,\delta}\bigr\rangle$
each $\tilde{H}_{k,\{i_{m}\}}$ must be characterized by higher accuracy
$(\epsilon',\mu)$, such that the number of measurements needed increases
to the value given in Eq. (\ref{eq:22 N'}). For a fixed $\alpha$
there are indeed \textit{infinite passive observables} whose corresponding
tests can be simultaneously evaluated. This is a consequence of the
two observations stated below. Importantly, we must keep in mind that
$\alpha$ is a positive integer and $\delta$ is a positive real number.
\begin{itemize}
\item For $\alpha=\alpha_{\textrm{max}}$, direct estimation yields the
set $\{\tilde{H}_{k,\{i_{m}\}}\}_{1\leq k\leq\alpha_{\textrm{max}},\{i_{m}\}}$,
which includes all the estimated values $\{\tilde{H}_{k,\{i_{m}\}}\}_{1\leq k\leq\alpha,\{i_{m}\}}$,
$\alpha\leq\alpha_{\textrm{max}}$. This allows to obtain all the
$\{\Delta F_{\alpha,\delta}\}_{\alpha\leq\alpha_{\textrm{max}}}$
for a certain value of $\delta$, using Eq. (\ref{eq:21.05 mean value of F in terms of <H_k,=00007Bim=00007D>}). 
\item Let us rewrite $N'$ in Eq. (\ref{eq:22 N'}) as $N'(\alpha,\delta)$,
to make explicit the dependence on $\alpha$ and $\delta$. Since
$N'(\alpha,\delta)\leq N'(\alpha_{\textrm{max}},\delta_{\textrm{max}})$
for $\alpha\leq\alpha_{\textrm{max}}$ and $\delta\leq\delta_{\textrm{max}}$,
$N'(\alpha_{\textrm{max}},\delta_{\textrm{max}})$ measurements suffice
to obtain any estimated value $\{\Delta F_{\alpha,\delta}\}_{\alpha\leq\alpha_{\textrm{max}},\delta\leq\delta_{\textrm{max}}}$,
with accuracy $(2\epsilon,(1-\mu)^{2})$, by applying Eq. (\ref{eq:21.05 mean value of F in terms of <H_k,=00007Bim=00007D>}). 
\end{itemize}
We also remark that the previous procedure is directly applicable
to the deformed observables $\mathcal{F}_{\alpha,\delta}^{(def)}$.
In this case the only difference is that Eq. (\ref{eq:21.05 mean value of F in terms of <H_k,=00007Bim=00007D>})
must be applied using coefficients $c_{k,\{i_{m}\}}^{(\alpha,\delta)}=\left(\begin{array}{c}
\alpha\\
k
\end{array}\right)(-\delta)^{\alpha-k}\prod_{m=1}^{k}\beta_{i_{m}}^{(def)}$, where $\{\beta_{i}^{(def)}\}$ are the effective temperatures that
characterize a given deformation. Importantly, if there are sufficiently
hot qubits in the initial state it is possible to use deformations
that set the corresponding temperatures to zero \citep{uzdin2018global},
i.e. $\beta_{i}\rightarrow\beta_{i}^{(def)}=0$. This is illustrated
in the second experiment of the main text with the inverse temperature
$\beta_{1}$. For such deformations, $\mathcal{B}$ is transformed
into $\mathcal{B}_{def}=\sum'_{i}\beta_{i}^{(def)}H_{i}$, where prime
in the sum indicates that it convers only a subset of qubits. If $n_{def}$
denotes the number of qubits in this subset, the substitution of $\mathcal{B}$
by $\mathcal{B}_{def}$ in Eq. (\ref{eq:20.1 passive observable})
leads to analogous of Eqs. (\ref{eq:21.1 total error}) and (\ref{eq:22 N'})
where $n$ is replaced by $n_{def}$. Accordingly, some deformations
can significantly reduce the measurement cost $N'$ if $n_{def}\ll n$. 

\section{sensitivity of heat leak tests using passive observables and practical
advantage }

In this appendix we characterize an infinite family of heat leaks
whose diagnostics is inefficient using population estimation, i.e.
it requires a number of measurements of order $\mathcal{O}(2^{n})$.
Moreover, by construction such heat leaks are detectable using passive
observables. This illustrates a situation where the efficiency of
the method can be fully exploited for actual detection. 

We start by analyzing the complexity of the method described in the
previous appendix, which provides detection if the inequality $\sum_{i=1}^{j}r_{i}^{\downarrow}<\sum_{i=1}^{j}s_{i}^{\downarrow}$
holds for some $1\leq j\leq2^{n}$. Since $\rho$ is diagonal in the
computational basis (cf. Eq. (1) of the main text), the most convenient
choice is $s_{i}^{\downarrow}=p'{}_{i}^{\downarrow}$, where $p'_{i}=\textrm{Tr}(|i_{\textrm{c}}\rangle\langle i_{\textrm{c}}|\rho)$
(for simplicity we also write $r_{i}^{\downarrow}=p_{i}^{\downarrow}$,
with $p_{i}=\textrm{Tr}(|i_{\textrm{c}}\rangle\langle i_{\textrm{c}}|\rho)$).
Accordingly, the following recipe could be applied to heat leak diagnostics:
\begin{enumerate}
\item Estimate the initial and final populations $\{p_{i}\}$ and $\{p'_{i}\}$. 
\item Sort them in non-increasing order. 
\item Evaluate the quantities $\xi_{j}\equiv\sum_{i=1}^{j}p{}_{i}^{\prime\downarrow}-\sum_{i=1}^{j}p{}_{i}^{\downarrow}$.
If $\xi_{j}>0$ for some $1\leq j\leq2^{n}$, there is a heat leak.
If $\xi_{j}\leq0$ for $1\leq j\leq2^{n}$, there can be a heat leak
that is undetectable using this method. This kind of heat leak corresponds
to a transformation such that $\rho'$ is not majorized by $\rho$,
yet the majorization relation $\rho\succ D(\rho')$ (which is equivalent
to $\xi_{j}\leq0$ for $1\leq j\leq2^{n}$ ) holds. 
\end{enumerate}
Step 1 requires to estimate the $M=2^{n}-1$ mean values $\bigl\langle|i_{\textrm{c}}\rangle\langle i_{\textrm{c}}|\bigr\rangle$
for the initial and final states. This number contributes to $N$
in Eq. (\ref{eq:20 N}) with a term that is linear in $n$. However,
the shadow norm for projectors $|i_{\textrm{c}}\rangle\langle i_{\textrm{c}}|$
is exponential in $n$, for both Pauli measurements and measurements
based on general Clifford unitaries \citep{huang2020predicting}.
This implies that population estimation is inefficient with the method
developed in \citep{huang2020predicting}. On the other hand, it has
recently been pointed out that a \textit{single} population can be
efficiently estimated employing a different technique \citep{paini2020estimating}.
The key obstacle is that the results presented in \citep{paini2020estimating}
apply to single observables, and step 1 refers to exponentially many
observables. Crucially, even for a single value of $j$ the quantity
$\xi_{j}$ can be unambiguosly evaluated only if all the populations
have been estimated. For example, suppose that one chooses randomly
a set of projectors $\{|i_{\textrm{c}}\rangle\langle i_{\textrm{c}}|\}_{i\in I}$,
where $I=\{i_{1},i_{2},...,i_{j}\}\subset\{1,2,...,2^{n}-1\}$ is
some arbitrary set of $j\ll2^{n}$ indices. From the corresponding
populations one can be certain that the sum $\sum_{i\in I}p'_{i}$
equals $\sum_{i=1}^{j}p{}_{i}^{\prime\downarrow}$, \textit{without
measuring any other population}, iff 
\[
1-\sum_{i\in I}p'_{i}\leq\textrm{min}_{i\in I}p'_{i}.
\]

When this inequality is satisfied, it can be combined with the relation
$\textrm{max}_{i\notin I}p'_{i}\leq\sum_{i\notin I}p'_{i}=1-\sum_{i\in I}p'_{i}$
to conclude that $\textrm{max}_{i\notin I}p'_{i}\leq\textrm{min}_{i\in I}p'_{i}$.
Therefore, $\sum_{i\in I}p'_{i}$ contains indeed the largest $j$
populations. Conversely, if $1-\sum_{i\in I}p'_{i}>\textrm{min}_{i\in I}p'_{i}$,
it is possible that $\textrm{max}_{i\notin I}p'_{i}>\textrm{min}_{i\in I}p'_{i}$,
and $\sum_{i\in I}p'_{i}\neq\sum_{i=1}^{j}p{}_{i}^{\prime\downarrow}$.
However, without prior information about $\rho'$ it is extremely
unlikely that the set $I$ corresponds to $\{p{}_{i}^{\prime\downarrow}\}_{1\leq i\leq j}$,
since this is just one among $\frac{\left(2^{n}-1\right)!}{\left(2^{n}-1-j\right)!j!}$
possible sets of $j$ projectors. This allows us to conclude that
the evaluation of $\xi_{j}$ is inefficient. 

\textbf{Sensitivity of heat leak tests using passive observables}.
Let us now characterize the class of heat leaks that can be diagnosed
by performing tests $\Delta\bigl\langle\mathcal{F}\bigr\rangle$.
Although we previously showed that this can be done efficiently, it
is important to understand the fundamental limits on sensitivity for
these tests. If $f_{i}$ denotes the eigenvalue of $\mathcal{F}$
corresponding to the eigenstate $|i_{\textrm{c}}\rangle$, by definition
\begin{equation}
\Delta\bigl\langle\mathcal{F}\bigr\rangle=\sum_{i=1}^{d}f_{i}\Delta p_{i},\label{eq:23 Delta<F>}
\end{equation}
where $\Delta p_{i}=\textrm{Tr}[|i_{\textrm{c}}\rangle\langle i_{\textrm{c}}|(\rho'-\rho)]$.
For our purpose it is covenient to write $\Delta\bigl\langle\mathcal{F}\bigr\rangle$
in a different manner. By defining $\Delta f_{k}\equiv f_{k+1}-f_{k}$,
we have that $f_{i}=f_{0}+\sum_{k=0}^{i-1}\Delta f_{k}$, where $f_{0}$
is just a reference value that won't appear in the final expression.
The substitution of $f_{i}$ by $f_{0}+\sum_{k=0}^{i-1}\Delta f_{k}$
in Eq. (\ref{eq:23 Delta<F>}) yields 
\begin{align*}
\Delta\bigl\langle\mathcal{F}\bigr\rangle & =\sum_{i=1}^{d}\left(\sum_{j=0}^{i-1}\Delta f_{j}\right)\Delta p_{i}+f_{0}\sum_{i=1}^{d}\Delta p_{i}\\
 & =\sum_{j=1}^{d-1}\Delta f_{j}\left(\sum_{i=j+1}^{d}\Delta p_{i}\right)\\
 & =-\sum_{j=1}^{d-1}\Delta f_{j}\bar{\xi}_{j},
\end{align*}
where $\bar{\xi}_{j}\equiv\Delta\sum_{i=1}^{j}p_{i}$. 

It is interesting to compare the quantity $\bar{\xi}_{j}$ with $\xi_{j}$,
previously introduced for heat leak diagnostics based on population
estimation. First, note that if we consider the sorting $p_{j+1}\leq p_{j}$,
$\Delta f_{j}\geq0$ by definition of passive observable. This does
not mean that in order to \textit{experimentally} estimate $\Delta\bigl\langle\mathcal{F}\bigr\rangle$
such a sorting needs to be done; it is mereley a convenient choice
to analyze which heat leaks can be detected through $\Delta\bigl\langle\mathcal{F}\bigr\rangle$.
With this convention, any term $\bar{\xi}_{j}>0$ reduces the value
of $\Delta\bigl\langle\mathcal{F}\bigr\rangle$ and viceversa. Hence,
$\Delta\bigl\langle\mathcal{F}\bigr\rangle<0$ only if $\bar{\xi}_{j}>0$
for at least one $1\leq j\leq2^{n}-1$.On the other hand, $\bar{\xi}_{j}>0$
implies $\xi_{j}>0$, since $\sum_{i=1}^{j}p'{}_{i}^{\downarrow}\geq\sum_{i=1}^{j}p'{}_{i}$.
Therefore, heat leak detection via $\Delta\bigl\langle\mathcal{F}\bigr\rangle$
implies detection via $\{\bar{\xi}_{j}\}$, which in turn implies
detection via $\{\xi_{j}\}$. However, the implementation of a set
of tests $\{\bar{\xi}_{j}\}_{1\leq j\leq J}$ is also inefficient
for values of $J$ such that $J\sim\mathcal{O}\left(2^{n}\right)$,
since it requires estimating an exponential amount of final populations.
This means that for $n$ large such tests can only provide efficient
diagnostics of heat leaks that affect ``small-size'' partial sums
$\sum_{i=1}^{j}p_{i}$, characterized by $j\ll2^{n}$. Note that the
cost of estimating the initial populations is not included in the
analysis. This stems from the assumption of initial product states
$\rho=\otimes_{k=1}^{n}\rho_{k}^{\beta_{k}}$, which can be efficiently
reconstructed using matrix product state tomography \citep{cramer2010efficient}. 

We have seen that the test $\Delta\bigl\langle\mathcal{F}\bigr\rangle$
is sensitive to heat leaks (i.e. $\Delta\bigl\langle\mathcal{F}\bigr\rangle<0$)
only if $\bar{\xi}_{j}<0$ for some $1\leq j\leq2^{n}-1$. Since for
$J\ll2^{n}$ it may be possible to efficiently evaluate the tests
$\{\bar{\xi}_{j}\}_{1\leq j\leq J}$, a question that naturally arises
is if $\Delta\bigl\langle\mathcal{F}\bigr\rangle$ is sensitive to
heat leaks that are \textit{outside the range of efficient detection
using} $\{\bar{\xi}_{j}\}$. If so, we can show a \textit{practical
advantage} of $\Delta\bigl\langle\mathcal{F}\bigr\rangle$ over the
direct evaluation of $\{\bar{\xi}_{j}\}$. To this end we introduce
the following theorem. Such a theorem allows us to characterize an
infinite family of heat leaks that are detectable using $\Delta\bigl\langle\mathcal{F}\bigr\rangle$
but elude efficient detection through $\{\bar{\xi}_{j}\}$. 

\textbf{Theorem 1}. Let $\rho\rightarrow\rho'$ be a transformation
such that $\Delta p_{i}\geq0$ for all $1\leq i\leq J$, $\Delta p_{i}>0$
for at least some $1\leq i\leq J$, and $\Delta p_{i}\leq0$ for all
$J+1\leq i\leq d$. For any non-trivial passive observable $\mathcal{F}$
it holds that $\Delta\bigl\langle\mathcal{F}\bigr\rangle=\textrm{Tr}[\mathcal{F}(\rho'-\rho)]<0$. 

\textbf{Proof}. Without loss of generality, we can assume that $f_{i}\geq0$
for $1\leq i\leq d$. In this way, the hypothesis of the theorem implies
$\sum_{i=1}^{J}f_{i}\Delta p_{i}\leq\textrm{max}_{i\leq J}f_{i}\sum_{i=1}^{J}\Delta p_{i}$,
and $\textrm{min}_{J+1\leq i}f_{i}\sum_{i=J+1}^{d}\left(-\Delta p_{i}\right)\leq\sum_{i=J+1}^{d}f_{i}\left(-\Delta p_{i}\right)$.
From probability conservation ($\sum_{i=1}^{J}\Delta p_{i}=\sum_{i=J+1}^{d}\left(-\Delta p_{i}\right)$)
and the monotonicity condition $f_{i}\leq f_{i+1}$ it also follows
that $\textrm{max}_{i\leq J}f_{i}\sum_{i=1}^{J}\Delta p_{i}\leq\textrm{min}_{J+1\leq i}f_{i}\sum_{i=J+1}^{d}\left(-\Delta p_{i}\right)$.
Hence, we can use this inequality to join the first two and obtain
$\sum_{i=1}^{J}f_{i}\Delta p_{i}\leq\sum_{i=J+1}^{d}f_{i}\left(-\Delta p_{i}\right)$,
which is equivalent to $\Delta\bigl\langle\mathcal{F}\bigr\rangle\leq0$
(see Eq. (\ref{eq:23 Delta<F>})). Since at least one of the employed
inequalities is strict unless $f_{i}=f_{i+1}$ for all $i$, $\Delta\bigl\langle\mathcal{F}\bigr\rangle<0$. 

Now we proceed to characterize a family of transformations $\{\rho\rightarrow\rho'=\mathcal{E}(\rho)\textrm{ }\textrm{s.t. }\mathcal{E}=\mathcal{E}_{\textrm{un}}\oplus\mathcal{E}_{\textrm{hl}}\}$,
where $\mathcal{E}$ is a direct sum of a unital map $\mathcal{E}_{\textrm{un}}$
and a (CPTP) map $\mathcal{E}_{\textrm{hl}}$ whose effect complies
with the hypothesis of Theorem 1. Specifically, $\mathcal{E}_{\textrm{hl}}$
acts non trivially on a subspace $\mathcal{H}_{\textrm{hl}}=\textrm{span}\{|i_{\textrm{c}}\rangle\}_{i\in I_{\textrm{hl}}}$,
and generates a heat leak characterized by the conditions: $\Delta p_{i}>0$
for $i\in I_{\textrm{hl}}$ and $i\leq J$, and $\Delta p_{i}<0$
for $i\in I_{\textrm{hl}}$ and $i\geq J+1$. In other words, if we
define the state $\rho_{\textrm{hl}}\equiv\sum_{i\in I_{\textrm{hl}}}\frac{p_{i}}{\sum_{i\in I_{\textrm{hl}}}p_{i}}|i_{\textrm{c}}\rangle\langle i_{\textrm{c}}|$,
the transformation $\rho_{\textrm{hl}}\rightarrow\rho'_{\textrm{hl}}=\mathcal{E}_{\textrm{hl}}(\rho_{\textrm{hl}})$
satisfies Theorem 1. Hence, $\textrm{Tr}[\mathcal{F}(\rho'_{\textrm{hl}}-\rho_{\textrm{hl}})]<0$.
The map $\mathcal{E}_{\textrm{un}}$ induces a transformation $\rho_{\textrm{un}}\rightarrow\rho'_{\textrm{un}}=\mathcal{E}_{\textrm{un}}(\rho_{\textrm{un}})$,
with $\rho_{\textrm{un}}\equiv\sum_{i\notin I_{\textrm{hl}}}\frac{p_{i}}{\sum_{i\notin I_{\textrm{hl}}}p_{i}}|i_{\textrm{c}}\rangle\langle i_{\textrm{c}}|$
being a state with support in the complement subspace of $\mathcal{H}_{\textrm{hl}}$,
denoted as $\mathcal{H}_{\textrm{un}}$. Noting that $\rho$ can be
written as $\rho=p_{\textrm{hl}}\rho_{\textrm{hl}}+(1-p_{\textrm{hl}})\rho_{\textrm{un}}$,
with $p_{\textrm{hl}}=\sum_{i\in I_{\textrm{hl}}}p_{i}$, it follows
that $\mathcal{E}(\rho)=p_{\textrm{hl}}\mathcal{E}_{\textrm{hl}}(\rho_{\textrm{hl}})+(1-p_{\textrm{hl}})\mathcal{E}_{\textrm{un}}(\rho_{\textrm{un}})$. 

Note that the transformations described in Theorem 1 can be recovered
from the more general transformations $\rho\rightarrow\mathcal{E}(\rho)$,
if we set $\mathcal{E}_{\textrm{un}}$ to be the identity $\mathbb{I}_{\textrm{un}}$
on $\mathcal{H}_{\textrm{un}}$. For the transformations $\rho\rightarrow\mathcal{E}(\rho)$
the test $\Delta_{\mathcal{E}}\bigl\langle\mathcal{F}\bigr\rangle\equiv\textrm{Tr}[\mathcal{F}(\mathcal{E}(\rho)-\rho)]$
contains a positive contribution $\Delta_{\textrm{un}}\bigl\langle\mathcal{F}\bigr\rangle\equiv\textrm{Tr}[\mathcal{F}(\rho'_{\textrm{un}}-\rho_{\textrm{un}})]$
and a negative contribution $\Delta_{\textrm{hl}}\bigl\langle\mathcal{F}\bigr\rangle\equiv\textrm{Tr}[\mathcal{F}(\rho'_{\textrm{hl}}-\rho_{\textrm{hl}})]$.
That is, 
\[
\Delta_{\mathcal{E}}\bigl\langle\mathcal{F}\bigr\rangle=p_{\textrm{hl}}\Delta_{\textrm{hl}}\bigl\langle\mathcal{F}\bigr\rangle+(1-p_{\textrm{hl}})\Delta_{\textrm{un}}\bigl\langle\mathcal{F}\bigr\rangle.
\]
Accordingly, 
\begin{equation}
\Delta_{\mathcal{E}}\bigl\langle\mathcal{F}\bigr\rangle<0\Leftrightarrow\frac{\bigl|\Delta_{\textrm{hl}}\bigl\langle\mathcal{F}\bigr\rangle\bigr|}{\Delta_{\textrm{un}}\bigl\langle\mathcal{F}\bigr\rangle}>\frac{1-p_{\textrm{hl}}}{p_{\textrm{hl}}}.\label{eq:24}
\end{equation}

We can now characterize heat leaks that satisfy Eq. (\ref{eq:24})
but do not admit efficient diagnostics through $\{\bar{\xi}_{j}\}$.
More formally, this means that $\bar{\xi}_{j}\geq0$ for $2^{n-1}-k\leq j\leq2^{n-1}+l$,
where $k,l\ll2^{n-1}$, and $\bar{\xi}_{j}\leq0$ otherwise. This
implies that it is necessary to estimate $\mathcal{O}(2^{n-1})$ populations
in order to detect the positivity of one of the quantities $\bar{\xi}_{j}$.
Importantly, for $j$ close to $2^{n}$ the corresponding $\bar{\xi}_{j}$
can in principle be efficiently evaluated as $\bar{\xi}_{j}=-\sum_{i=j+1}^{2^{n}}\Delta p_{i}$,
which is why we consider also $j\leq2^{n-1}+l$. Transformations $\rho\rightarrow\rho'=\mathcal{E}(\rho)$
such that $\textrm{min}\left(I_{\textrm{hl}}\right)\geq2^{n-1}-k$
and $\textrm{max}\left(I_{\textrm{hl}}\right)\leq2^{n-1}+l$ satisfy
these conditions. Since by construction they are unital on $\mathcal{H}_{\textrm{un}}=\textrm{span}\{|i_{\textrm{c}}\rangle\}_{i\notin I_{\textrm{hl}}}$,
$\bar{\xi}_{j}>0$ for $j\leq2^{n-1}-k-1$ and $j\geq2^{n-1}+l+1$.
Moreover, there are infinite transformations of this kind that also
satisfy Eq. (\ref{eq:24}) for a given initial state $\rho$, as stated
below:
\begin{itemize}
\item All the subsets of indices $I_{\textrm{hl}}\subseteq\{i\textrm{ s.t. }2^{n-1}-k\leq i\leq2^{n-1}+l\}$
guarantee that heat leaks associated with transformations $\rho\rightarrow\rho'=\mathcal{E}(\rho)$
cannot be efficiently detected using $\{\bar{\xi}_{j}\}$. For any
subset $I_{\textrm{hl}}$ of this form, there are infinite transformations
$\rho_{\textrm{hl}}\rightarrow\rho'_{\textrm{hl}}$ corresponding
to infinite choices of population variations $\{\Delta p_{i}\}_{\in I_{\textrm{hl}}}$,
such that they satisfy the strict inequalities in Theorem 1 (i.e.
$\Delta p_{i}>0$ or $\Delta p_{i}<0$). 
\item Similarly, for any subset $I_{\textrm{hl}}\subseteq\{i\textrm{ s.t. }2^{n-1}-k\leq i\leq2^{n-1}+l\}$
there are infinite transformations $\rho_{\textrm{un}}\rightarrow\rho'_{\textrm{un}}$
that fulfill unitality, which implies that $\bar{\xi}_{j}>0$ for
$j\leq2^{n-1}-k-1$ and $j\geq2^{n-1}+l+1$. The chosen subset $I_{\textrm{hl}}$
determines the probability $p_{\textrm{hl}}=\sum_{i\in I_{\textrm{hl}}}p_{i}$
in Eq. (\ref{eq:24}) and the population changes $\{\Delta p_{i}\}_{\in I_{\textrm{hl}}}$
determine $\bigl|\Delta_{\textrm{hl}}\bigl\langle\mathcal{F}\bigr\rangle\bigr|$.
Given these parameters there are infinite transformations $\rho_{\textrm{un}}\rightarrow\rho'_{\textrm{un}}$
such that $\Delta_{\textrm{un}}\bigl\langle\mathcal{F}\bigr\rangle$
fulfills Eq. (\ref{eq:24}). This can be seen by considering first
$\mathcal{E}_{\textrm{un}}=\mathbb{I}_{\textrm{un}}$, which yields
$\Delta_{\textrm{un}}\bigl\langle\mathcal{F}\bigr\rangle=0$, and
then looking at the continuum of unital maps $\mathcal{E}_{\textrm{un}}$
that are not ``too far from $\mathbb{I}_{\textrm{un}}$'' to produce
a change $\Delta_{\textrm{un}}\bigl\langle\mathcal{F}\bigr\rangle>0$
that violates Eq. (\ref{eq:24}). 
\end{itemize}
To conclude this appendix, we stress that the results presented here
constitute \textit{sufficient} conditions for heat leak detection
using passive observables, which also guarantee that detection by
other methods is inefficient. However, it is very possible that such
an advantage also holds under more general circumstances. For example,
it is expected that for transformations fulfilling Eq. (\ref{eq:24})
small deviations from the condition $\mathcal{E}=\mathcal{E}_{\textrm{un}}\oplus\mathcal{E}_{\textrm{hl}}$
still adhere to it. In addition, $\mathcal{E}_{\textrm{hl}}$ is constructed
in such a way that it generates a heat leak detectable by \textit{any}
passive observable. Clearly, Eq. (\ref{eq:24}) is independent of
this assumption and only requires that $\Delta_{\textrm{hl}}\bigl\langle\mathcal{F}\bigr\rangle<0$
for a \textit{particular} observable $\mathcal{F}$. This indicates
that even though there are infinite heat leaks that satisfy Eq. (\ref{eq:24}),
the total set that does not admit efficient detection using other
techniques may be much larger. 

\section{\textit{\emph{Detector noise and characterization of initial states }}}

The measurement error is modeled as follows. For a general state $\rho$,
resulting from the application of some circuit to the ground state,
let $\{p_{i}\}$ be the ideal populations in the computational basis,
i.e. the populations that would be obtained if the detectors were
error-free. Experimentally, there is a finite probability $p(j|i)$
that the state registered by the detector is $|j\rangle$, given that
the projected state (i.e. the state corresponding to the ideal measurement)
is $|i\rangle$. The conditional probabilities $p(j|i)$ thus encapsulate
the effect of the detector noise, and yield the total probability
\begin{equation}
q_{j}=\sum_{i}p(j|i)p_{i},\label{eq:7}
\end{equation}
with $p(j|i)=\delta_{j,i}$ for ideal detectors. This gives rise to
a measurement matrix 
\begin{equation}
\mathbb{M}\equiv\left(\begin{array}{ccccc}
p(0|0) & p(0|1) & \cdots & p(0|i) & \cdots\\
p(1|0) & p(1|1)\\
\vdots &  & \ddots\\
p(j|0) &  &  & p(j|i)\\
\vdots &  &  &  & \ddots
\end{array}\right),\label{eq:8}
\end{equation}
which relates the vectors of ideal and experimental populations through
the equality $\{q_{i}\}=\mathbb{M}\{p_{i}\}$.

The $i$th column of the measurement matrix is experimentally determined
by preparing and measuring the state $|i\rangle$ of the computational
basis. Once $\mathbb{M}$ is constructed, the vector $\mathbb{M}^{-1}\{q_{i}\}$
(where $\mathbb{M}^{-1}$ is the inverse of $\mathbb{M}$) provides
an estimation of the ideal populations $\{p_{i}\}$. In the case of
the experiments implemented in the Melbourne processor we compute
independently measurement matrices for the system and for the environment,
by running the corresponding computational bases. From the initial
system populations $\{p_{i}^{s}\}$, we determine the closest state
of the form $\otimes_{k=9}^{12}\rho_{k}$, where each $\rho_{k}$
is a diagonal qubit state. Specifically, we numerically evaluate the
minimum $\textrm{min}_{\otimes_{k=9}^{12}\rho_{k}}\left\Vert \{p_{i}^{s}\}-\otimes_{k=9}^{12}\rho_{k}\right\Vert _{2}$,
being $\left\Vert \cdot\right\Vert _{2}$ the L-2 norm. The state
that yields the minimum is characterized by ground (qubit) populations
$p_{0}^{(12)}=0.612$, $p_{0}^{(11)}=0.586$, $p_{0}^{(10)}=0.611$,
and $p_{0}^{(9)}=0.557$, and the minimized value itself is $0.005$.
In addition, the inverse measurement matrix of the environment qubit
yields ground population $p_{0}^{(8)}=p_{0}^{e}=0.782$. The measurement
matrix associated to the Essex processor is constructed by running
the computational basis of the total system, including the three-qubit
system and the environment. After removing the detector noise through
the application of $\mathbb{M}^{-1}$, the ground qubit populations
for the system and the environment are given respectively by $\{p_{0}^{(0)},p_{0}^{(1)},p_{0}^{(3)}\}=\{0.944,0.652,0.652\}$
and $p_{0}^{(4)}=p_{0}^{e}=0.806$. 

According to the previous results, we note that for both processors
the measured initial state can be reliably described by Eq. (1) of
the main text. On the other hand, in the case of the Melbourne processor
there is an important difference between this state and the theoretical
initial state, which is coded in the software interface of the IBM
quantum experience platform. The coded system state has ground populations
$p_{0}^{(12)}=p_{0}^{(11)}=p_{0}^{(10)}=0.578$ and $p_{0}^{(9)}=0.654$,
and the coded environment state has ground population $p_{0}^{(4)}=0.875$.
This is in stark contrast with the Essex processor, where the coded
system state and coded environment state are respectively characterized
by populations $\{p_{0}^{(0)},p_{0}^{(1)},p_{0}^{(3)}\}=\{0.945,0.654,0.654\}$
and $p_{0}^{(4)}=0.793$. We attribute such a difference to the employment
of cnot gates for the preparation performed in the Melbourne processor,
which are noisier than single-qubit gates. To overcome this technical
limitation, the heat leak tests performed with this processor are
based on passive observables constructed from the measured initial
state. 

\section{Statistical error }

The uncertainty of a heat leak test quantifies the fluctuations in
the value of $\Delta\left\langle \mathcal{F}\right\rangle $ for different
repetitions of the same experiment. A single experiment refers to
the implementation of two independent circuits for the initial and
final states, each of which is sampled by performing a certain number
$N$ of single-shot measurements. In this way, the initial and final
mean values $\left\langle \mathcal{F}\right\rangle _{0}$ and $\left\langle \mathcal{F}\right\rangle _{f}$
are computed using the results of $N$ shots, and $\Delta\left\langle \mathcal{F}\right\rangle =\left\langle \mathcal{F}\right\rangle _{f}-\left\langle \mathcal{F}\right\rangle _{0}$.
The calculation of the theoretical uncertainty is simplified by taking
into account that, by construction, the initial and final distributions
for the eigenvalues of $\mathcal{F}$ are independent. If $p_{i,j}$
denotes the probability to measure $\mathcal{F}_{i}$ (where $\mathcal{F}_{i}$
is an eigenvalue of $\mathcal{F}$) for the initial state and $\mathcal{F}_{j}$
for the final state, then $p_{i,j}=p_{i}p'_{j}$, being $p_{i}$ the
initial probability to measure $\mathcal{F}_{i}$ and $p'_{j}$ the
final probability to measure $\mathcal{F}_{j}$. In this way, the
variance for a single measurement of $\mathcal{F}$ at the beginning
and at the end is given by 
\begin{align}
\textrm{Var}_{\textrm{shot}}(\Delta\mathcal{F}) & =\sum_{i,j}p_{i,j}\left((\mathcal{F}_{j}-\mathcal{F}_{i})-\left\langle \Delta\mathcal{F}\right\rangle \right)^{2}\nonumber \\
 & =\left\langle \left(\Delta\mathcal{F}\right)^{2}\right\rangle -\left\langle \Delta\mathcal{F}\right\rangle ^{2}\nonumber \\
 & =\textrm{Var}_{\textrm{shot}}(\mathcal{F})_{0}+\textrm{Var}_{\textrm{shot}}(\mathcal{F})_{f},\label{eq:9}
\end{align}
where $\left\langle \left(\Delta\mathcal{F}\right)^{2}\right\rangle =\sum_{i,j}p_{i,j}(\mathcal{F}_{j}-\mathcal{F}_{i})^{2}$
and $\left\langle \Delta\mathcal{F}\right\rangle =\sum_{i,j}p_{i,j}(\mathcal{F}_{j}-\mathcal{F}_{i})=\Delta\left\langle \mathcal{F}\right\rangle $.
The expression in the third line is the sum of the initial variance
$\textrm{Var}(\mathcal{F})_{0}=\sum_{i}p_{i}$$\left(\mathcal{F}_{i}-\left\langle \mathcal{F}\right\rangle _{0}\right)^{2}$
and the final variance $\textrm{Var}(\mathcal{F})_{f}=\sum_{j}p'_{j}$$\left(\mathcal{F}_{j}-\left\langle \mathcal{F}\right\rangle _{f}\right)^{2}$. 

From the central limit theorem, if $N$ measurements of $\mathcal{F}$
are performed at the beginning and at the end, the variances for the
corresponding mean values are the single-shot variances reduced by
a factor of $1/N$. Therefore, the theoretical variance predicted
for a single experiment is 
\begin{align}
\textrm{Var}(\Delta\mathcal{F}) & =\textrm{Var}(\mathcal{F})_{0}+\textrm{Var}(\mathcal{F})_{f}\nonumber \\
 & =\frac{1}{N}\left[\textrm{Var}_{\textrm{shot}}(\mathcal{F})_{0}+\textrm{Var}_{\textrm{shot}}(\mathcal{F})_{f}\right].\label{eq:12}
\end{align}
The value of $N$ for the experiments with the Melbourne processor
is $N_{\textrm{Mel}}=8192$, i.e. the number of shots for each preparation
and evolution batch. For the experiments realised with the Essex processor
$N_{\textrm{Ess}}=16\times8192$, with the factor $16$ accounting
for the sixteen coherent states involved in the preparation of the
initial state. 

On the other hand, the experimental variances $\textrm{Var}_{\textrm{exp}}(\mathcal{F})_{0}$
and $\textrm{Var}_{\textrm{exp}}(\mathcal{F})_{f}$ are computed as
\begin{align}
\textrm{Var}_{\textrm{exp}}\left(\mathcal{F}\right)_{0} & =\frac{1}{n-1}\sum_{k=1}^{n}\left(\bar{\mathcal{F}}_{k}-\frac{1}{n}\sum_{k=1}^{n}\bar{\mathcal{F}}_{k}\right)^{2},\label{eq:13}\\
\textrm{Var}_{\textrm{exp}}\left(\mathcal{F}\right)_{f} & =\frac{1}{n-1}\sum_{k=1}^{n}\left(\bar{\mathcal{F}}'_{k}-\frac{1}{n}\sum_{k=1}^{n}\bar{\mathcal{F}}'_{k}\right)^{2},\label{eq:14}
\end{align}
where $\bar{\mathcal{F}}_{k}$ and $\bar{\mathcal{F}}'_{k}$ are respectively
the (experimental) initial and final mean values of $\mathcal{F}$
corresponding to the $k$th batch. For the Melbourne processor, there
are $n=n_{\textrm{Mel}}=10$ batches of 8192 shots each. For the Essex
processor, there are $n=n_{\textrm{Mel}}=4$ batches of $16\times8192$
shots each. The total experimental variance is 
\begin{equation}
\textrm{Var}_{\textrm{exp}}(\Delta\mathcal{F})=\textrm{Var}_{\textrm{exp}}\left(\mathcal{F}\right)_{0}+\textrm{Var}_{\textrm{exp}}\left(\mathcal{F}\right)_{f}.\label{eq:15}
\end{equation}
The confidence intervals in the plots of the main text are given by
three standard deviations above and below the mean value of $\Delta\left\langle \mathcal{F}\right\rangle $
over all the experiments, with the theoretical and experimental standard
deviations computed by taking the square root of Eqs. (\ref{eq:12})
and (\ref{eq:15}), respectively. 

\section{Preparation of a product of thermal states by mixing coherent states }

Consider a mixture of two coherent states of a qubit,

\begin{equation}
\rho=\frac{1}{2}\left(R_{y}(\theta)|0\rangle\langle0|R_{y}^{\dagger}(\theta)+R_{y}(-\theta)|0\rangle\langle0|R_{y}^{\dagger}(-\theta)\right),\label{eq:16}
\end{equation}
where $R_{y}(\theta)$ is a rotation of $\theta$ degrees around the
$y$ axis in the Bloch sphere. While the states $R_{y}(\theta)|0\rangle$
and $R_{y}(-\theta)|0\rangle$ have coherence in the energy basis
(defined by the igenstates $\{|0\rangle,|1\rangle\}$), the mixture
(\ref{eq:16}) is the diagonal state 
\begin{equation}
\rho=\textrm{cos}^{2}\left(\frac{\theta}{2}\right)|0\rangle\langle0|+\textrm{sin}^{2}\left(\frac{\theta}{2}\right)|1\rangle\langle1|,\label{eq:17}
\end{equation}
which represents a thermal state for $\theta\leq\frac{\pi}{2}$. This
is readily deduced by substituting the explicit expressions 
\begin{equation}
R_{y}(\pm\theta)|0\rangle=\textrm{cos}\left(\frac{\theta}{2}\right)|0\rangle\pm\textrm{sin}\left(\frac{\theta}{2}\right)|1\rangle,\label{eq:18}
\end{equation}
 into Eq. (\ref{eq:16}). 

A product of an arbitrary number $N$ of thermal qubit states can
also be expressed as a mixture analogous to Eq. (\ref{eq:16}). Let
$\rho_{i}(\theta_{i})=\textrm{cos}^{2}\left(\frac{\theta_{i}}{2}\right)|0\rangle_{i}\langle0|+\textrm{sin}^{2}\left(\frac{\theta_{i}}{2}\right)|1\rangle_{i}\langle1|$
be the state of the $i$th qubit, and let $|\psi(\theta)\rangle_{i}\equiv R_{y}(\theta)|0\rangle_{i}$.
By writing each $\rho_{i}(\theta_{i})$ as in Eq. (\ref{eq:16}),
we obtain 
\begin{align}
\otimes_{i=1}^{N}\rho_{i}(\theta_{i}) & =\otimes_{i=1}^{N}\sum_{\theta=\pm\theta_{i}}\frac{1}{2}\left(|\psi(\theta)\rangle_{i}\langle\psi(\theta)|\right)\nonumber \\
 & =\frac{1}{2^{N}}\left[\sum_{\boldsymbol{\theta}}\otimes_{i=1}^{N}|\psi(\theta_{i})\rangle_{i}\langle\psi(\theta_{i})|\right],\label{eq:19}
\end{align}
where $\boldsymbol{\theta}=(\theta_{1},\theta_{2},...,\theta_{N})$
is a vector that contains the rotation angles of all qubits, and the
sum $\sum_{\boldsymbol{\theta}}$ runs over the $2^{N}$ combinations
$(\pm\theta_{1},\pm\theta_{2},...,\pm\theta_{N})$ involving $\pm\theta_{i}$
rotations. The preparation in the Essex processor is performed by
following this method. Each coherent state $\otimes_{i=1}^{4}|\psi(\pm\theta_{i})\rangle_{i}$
is prepared by applying $R_{y}(\pm\theta_{i})$ rotations to the ground
state of each qubit, which results in a total of $2^{4}=16$ coherent
states. In this way, the product $\otimes_{i=1}^{4}\rho_{i}(\theta_{i})$
is obtained by asigning the same weight $\frac{1}{16}$ to all the
coherent states, which are then mixed according to Eq. (\ref{eq:19}). 

\section{Additional heat leak tests }

In this appendix we show the results of additional heat leak tests,
performed with the same experimental data used in the main text. Figure
S1(a) shows the result of the test $\Delta\left\langle \mathcal{B}^{\alpha}\right\rangle $,
for $0\leq\alpha\leq4$, applied to the experiments with the Melbourne
processor. The mean value of $\Delta\left\langle \mathcal{B}^{\alpha}\right\rangle $
is depicted by the red curve, and the upper and lower black curves
are obtained by adding and subtracting three standard deviations to
the red curve, respectively. Thus, the confidence interval is contained
within the black curves. We can see that while on average $\Delta\left\langle \mathcal{B}^{\alpha}\right\rangle $
becomes negative for $\alpha\apprge3.4$, the confidence interval
always contains positive values. Therefore, the test does not provide
unambiguous detection. In Fig. S1(b) the test $\Delta\left\langle \mathcal{F}_{2,\delta}\right\rangle $
is depicted. Importantly, for even powers $\alpha$ the shift $\delta$
must be restricted to guarantee the passivity of $\mathcal{F}_{2,\delta}$,
and the interval in Fig. S1(b) is chosen accordingly. Similarly to
the previous case, on average $\Delta\left\langle \mathcal{F}_{2,\delta}\right\rangle $
becomes negative for $\alpha\apprge1.6$, but even for $\alpha=2$
the confidence interval still contains positive values. 

\begin{figure}
\centering{}\includegraphics[scale=0.36]{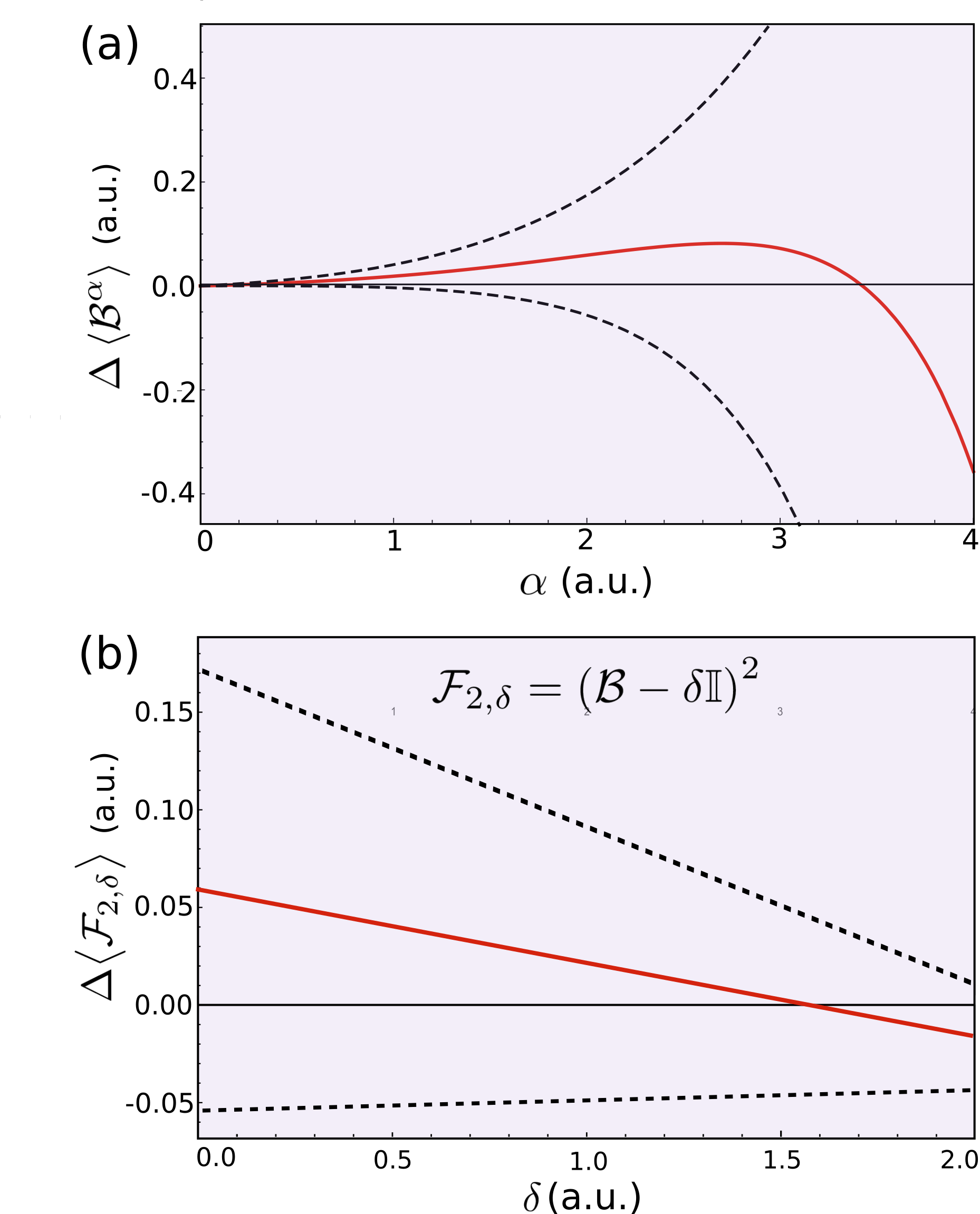}\caption{Additional heat leak tests using the experimental data of the Melbourne
processor. None of these tests yields unambiguous detection of the
heat leak.}
\end{figure}

Fig. S2 depicts heat leak tests corresponding to the Essex processor.
Figure S2(a) shows the result of the test $\Delta\left\langle \mathcal{F}_{5,\delta}^{(def)}\right\rangle $,
when the environment is decoupled. The observable $\mathcal{F}_{5,\delta}^{(def)}$
is given by $\mathcal{F}_{5,\delta}^{(def)}=(\mathcal{B}_{def}-\delta\mathbb{I})^{5}$,
with $\mathcal{B}_{def}$ the deformed observable indicated in the
main text. Consistently with the decoupling of the environment, $\Delta\left\langle \mathcal{F}_{5,\delta}^{(def)}\right\rangle \geq0$.
Finally, Fig. S2(b) shows that the test $\Delta\left\langle \mathcal{B}_{def}^{\alpha}\right\rangle $
yields unambiguous detection of the heat leak (when the environment
is coupled) for $\alpha\gtrsim6.8$. However, it is worth stressing
that the employement of the shift $\delta$ enables detection with
the smaller power $\alpha=5$, as shown in the main text. The inset
indicates that if the deformation is not a applied to $\mathcal{B}$,
no detection is possible for any $0\leq\alpha\leq7$.
\begin{figure}
\centering{}\includegraphics[scale=0.58]{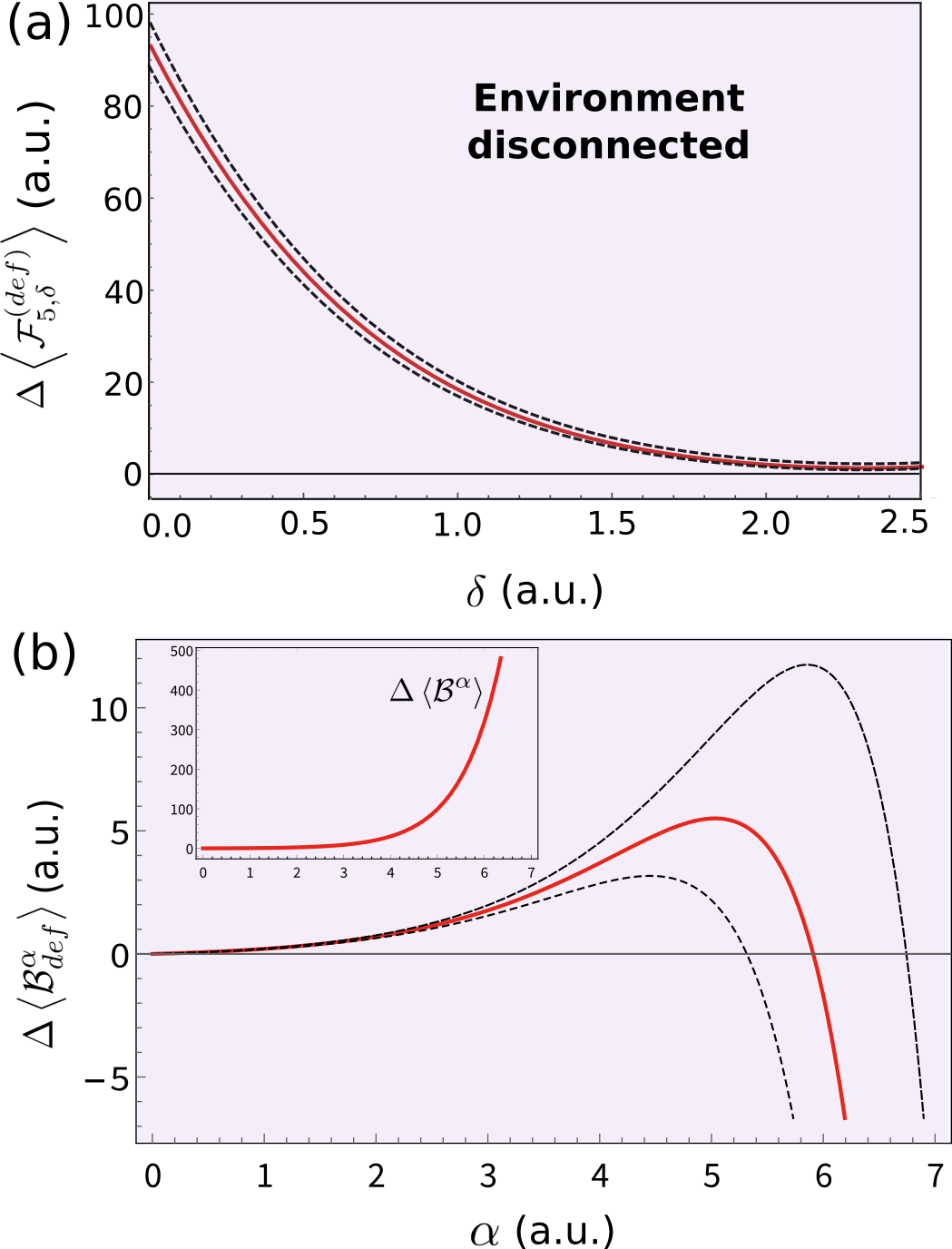}\caption{Additional heat leak tests using the experimental data of the Essex
processor. The test in (a) is performed for the case when the environment
is decoupled, and consistently yields only positive values. For the
experiments with the environment coupled, the test in (b) succesfully
detects the heat leak.}
\end{figure}

\section{Comparison between environment detection using resource theory and
global passivity}

The framework of thermodynamic resource theory (RT) is characterized
by a set of inequalities that govern the behavior of microscopic systems
under thermal operations \citep{Brandao2015}. Specifically, these
inequalities apply to transformations of the form 
\begin{equation}
\rho_{s}\otimes\sigma_{c}\rightarrow\rho'_{s}\otimes\sigma_{c}=\textrm{Tr}_{rc}U(\rho_{s}\otimes\sigma_{c}\otimes\rho_{r}^{\beta})U^{\dagger},\label{eq:28}
\end{equation}
where $\rho_{s}$ is the state of the system, $\rho_{r}^{\beta}$
is the state of a thermal bath (of arbitrary size) at inverse temperature
$\beta$, $\sigma_{c}$ is the state of a catalyst, and $U$ is an
energy-preserving unitary that acts globally in the aforementioned
systems. Energy conservation is characterized by the condition 
\begin{equation}
[U,H_{s}+H_{r}+H_{c}]=0,\label{eq:29}
\end{equation}
where $H_{s}$, $H_{r}$, and $H_{c}$ are respectively the Hamiltonians
of the system, the bath, and the catalyst. 

The RT inequalities constitute necessary and sufficient conditions
on the final state of the system, $\rho'_{s}$, in the case where
both $\rho_{s}$ and $\rho'_{s}$ commute with $H_{s}$. That is,
when $\rho_{s}=\sum_{i}p_{i}^{s}|i\rangle_{s}\langle i|$ and $\rho'_{s}=\sum_{i}p'{}_{i}^{s}|i\rangle_{s}\langle i|$,
being $|i\rangle_{s}$ eigenstates of $H_{s}$. If $q_{i}^{s}$ denote
the eigenvalues of the thermal state $\rho_{s}^{\beta}=\frac{e^{-\beta H_{s}}}{Z_{s}}$,
for any transformation obeying Eq. (\ref{eq:28}) none of the ``$\alpha$-free
energies'' 
\begin{equation}
F_{\alpha}(\rho_{s}||\rho_{s}^{\beta})=\beta^{-1}[D_{\alpha}(\rho_{s}||\rho_{s}^{\beta})-\textrm{ln}(Z_{s})],\:-\infty<\alpha<\infty,\label{eq:30}
\end{equation}
can increase. The quantity $D_{\alpha}(\rho_{s}||\rho_{s}^{\beta})$
is the ``$\alpha$-Renyi divergence'', defined as 
\begin{equation}
D_{\alpha}(\rho_{s}||\rho_{s}^{\beta})\equiv\frac{\textrm{sgn}(\alpha)}{\alpha-1}\textrm{ln}\left(\sum_{i}\left(p_{i}^{s}\right)^{\alpha}\left(q_{i}^{s}\right)^{1-\alpha}\right).\label{eq:31}
\end{equation}
In addition, if $F_{\alpha}(\rho'_{s}||\rho_{s}^{\beta})\leq F_{\alpha}(\rho_{s}||\rho_{s}^{\beta})$
for all $\alpha$ then there exist $\sigma_{c}$, $\rho_{r}^{\beta}$
and $U$ such that Eq. (\ref{eq:28}) holds \citep{Brandao2015}.

Consider the initial four-qubit state prepared in the Melbourne processor.
As explained in Section S-I, the closest product of thermal states
$\otimes_{i=9}^{12}\rho_{i}$ is characterized by ground qubit populations
$p_{0}^{(9)}=0.557$, $p_{0}^{(10)}=0.611$, $p_{0}^{(11)}=0.586$,
and $p_{0}^{(12)}=0.612\sim p_{0}^{(10)}$. In this way, we can have
several decompositions of the form 
\begin{equation}
\otimes_{i=9}^{12}\rho_{i}=\rho_{s}\otimes\rho_{r}^{\beta},\label{eq:32}
\end{equation}
where $\rho_{r}^{\beta}$ is a thermal state with possible inverse
temperatures $\beta=-\textrm{ln}\left(\frac{1-p_{0}^{(i)}}{p_{0}^{(i)}}\right)$,
$i=9,10,11$, and $\rho_{s}$ is the state of the remaining qubits.
Specifically, the role of the bath can be taken by any of the qubits,
or by the bipartite system formed by qubits 10 and 12, which share
the same temperature. The possible decompositions are thus
\begin{equation}
\otimes_{i=9}^{12}\rho_{i}=\rho_{s}\otimes\rho_{k},\label{eq:33}
\end{equation}
for $9\leq k\leq12$, and 
\begin{equation}
\otimes_{i=9}^{12}\rho_{i}=\rho_{s}\otimes(\rho_{10}\otimes\rho_{12}),\label{eq:34}
\end{equation}
with the state $\rho_{10}\otimes\rho_{12}$ characterized by the inverse
temperature $\beta=-\textrm{ln}\left(\frac{1-p_{0}^{(10)}}{p_{0}^{(10)}}\right)$. 

If $\otimes_{i=9}^{12}\rho_{i}$ evolves under a global energy-preserving
unitary, any system in the decompositions (\ref{eq:33}) and (\ref{eq:34})
must satisfy the RT inequalities, i.e. $F_{\alpha}(\rho'_{s}||\rho_{s}^{\beta})\leq F_{\alpha}(\rho_{s}||\rho_{s}^{\beta})$
for all $\alpha$. An example of such a unitary is provided by the
circuit in Fig. S3(a), consisting of only swap gates (here $H_{i}=|1\rangle_{i}\langle1|$
for $9\leq i\leq12$ and therefore each swap is energy-preserving).
On the other hand, suppose that first a swap takes place between an
external qubit $e$, prepared in a state $\rho_{e}=\rho_{11}$ (i.e.
with ground population $p_{0}^{(e)}=p_{0}^{(11)}$), and the qubit
9, as illustrated in Fig. S3(a). Since clearly this induces a non-unitary
dynamics on the four-qubit system, our goal is to determine if such
an interference can be detected by a violation of the form $F_{\alpha}(\rho'_{s}||\rho_{s}^{\beta})>F_{\alpha}(\rho_{s}||\rho_{s}^{\beta})$,
for some value of $\alpha$. By looking at the total final state $\otimes_{i=9}^{12}\rho'_{i}$
we deduce that such detection is impossible. The key observation is
that for any system in the decompositions (\ref{eq:33}) or (\ref{eq:34}),
the final state $\rho'_{s}$ is consistent with a transformation of
the form (\ref{eq:28}). Accordingly, all the $\alpha$-free energies
must decrease or remain unchanged. 
\begin{figure}
\centering{}\includegraphics[scale=0.62]{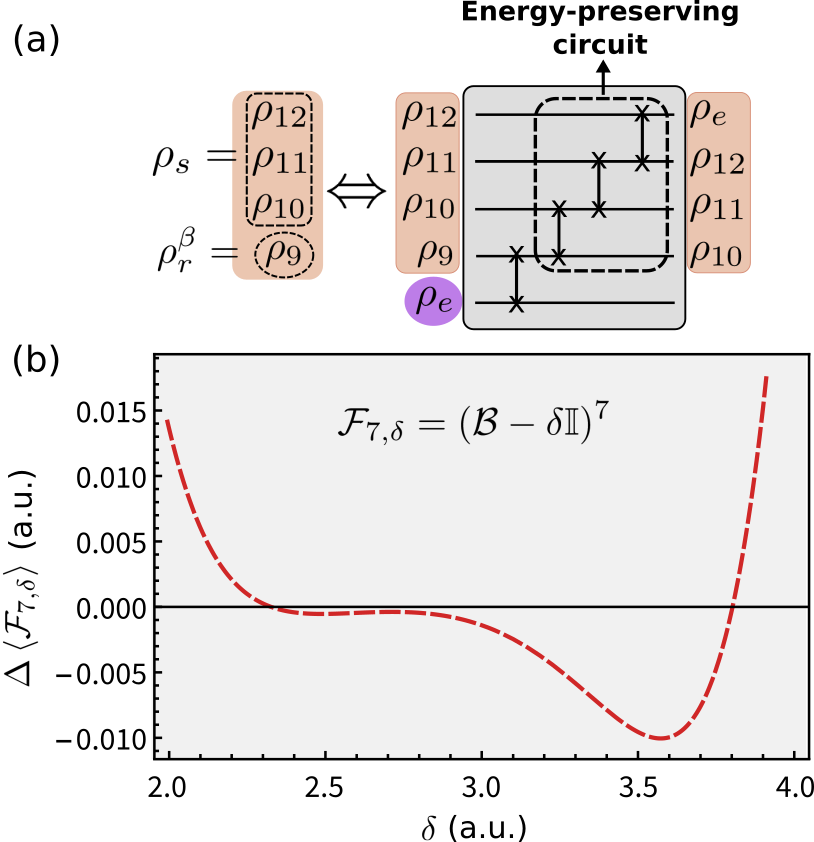}\caption{(a) A circuit used to check if a thermodynamic inequality of resource
theory is violated due to the coupling with the environment $\rho_{e}$.
The initial state $\otimes_{i=9}^{12}\rho{}_{i}$ of the total system
can be decomposed into a ``bath'', and a subsystem that should transform
obeying the resource theory inequalities if the environment were not
present. The figure illustrates a possible decomposition with the
qubit 9 as bath. (b) Heat leak test $\Delta\left\langle \mathcal{F}_{7,\delta}\right\rangle $
applied to the total system. Detection of the environment is observed
for $2.3\lesssim\delta\lesssim3.8$.}
\end{figure}

The total circuit in Fig. S3(a) transforms the qubits 9-12 into the
final state
\begin{equation}
\otimes_{i=9}^{12}\rho'_{i}=\rho_{10}\otimes\rho_{11}\otimes\rho_{10}\otimes\rho_{11},\label{eq:35}
\end{equation}
which means that the final state for qubits 9 and 11 is $\rho_{10}$
and the final state for qubits 10 and 12 is $\rho_{11}$. For the
system defined through Eqs. (\ref{eq:33}) and (\ref{eq:34}), the
final system state corresponding to Eq. (\ref{eq:35}) can also be
obtained without the interference of the environment. Specifically,
$\rho'_{s}$ can be generated by applying suitable combinations of
swaps and partial swaps between the qubits 9-12, on the initial state
$\otimes_{i=9}^{12}\rho$. Given that these operations satisfy Eq.
(\ref{eq:29}) (with the identity map applied to a potential catalyst),
all the RT inequalities must hold and therefore the environment cannot
be detected. The operations are explicitly the following: 
\begin{itemize}
\item If qubit 9 is the bath, Eq. (\ref{eq:35}) implies that $\rho_{s}=\rho_{10}\otimes\rho_{11}\otimes\rho_{12}$
is transformed into $\rho'_{s}=\rho_{11}\otimes\rho_{10}\otimes\rho_{11}$.
The transformation $\rho_{10}\otimes\rho_{11}\rightarrow\rho_{11}\otimes\rho_{10}$,
undergone by the qubits 10 and 11, is simply a total swap between
them. Moreover, the condition $p_{0}^{(9)}<p_{0}^{(11)}<p_{0}^{(12)}$
guarantees that the transformation $\rho_{12}\rightarrow\rho_{11}$
is possible through a partial swap between the qubit 12 and the qubit
9. 
\item If qubit 10 is the bath, the system transforms as $\rho_{s}=\rho_{9}\otimes\rho_{11}\otimes\rho_{12}\rightarrow\rho'_{s}=\rho_{10}\otimes\rho_{10}\otimes\rho_{11}$.
The final state of the qubit 9 can be achieved by swapping it with
the qubit 10. In addition, the equality $p_{0}^{(12)}=p_{0}^{(10)}$
implies that a swap between the qubits 11 and 12 yields the state
$\rho_{12}\otimes\rho_{11}=\rho_{10}\otimes\rho_{11}$. 
\item If qubit 11 is the bath, the system transforms as $\rho_{s}=\rho_{9}\otimes\rho_{10}\otimes\rho_{12}\rightarrow\rho'_{s}=\rho_{10}\otimes\rho_{11}\otimes\rho_{11}$.
This state can be achieved in two steps. First, a swap between the
qubits 9 and 10 yields $\rho_{10}\otimes\rho_{9}\otimes\rho_{12}$.
Since now the qubit 10 has ground population $p_{0}^{(9)}$, it is
not difficult to check that a suitable partial swap with the qubit
12 brings both qubits to the state $\rho_{11}$, thus completing the
transformation. 
\item If qubits 10 and 12 are the bath, the system transformation $\rho_{s}=\rho_{9}\otimes\rho_{11}\rightarrow\rho'_{s}=\rho_{10}\otimes\rho_{10}$
is achieved by simply swapping the system and the bath (recall that
$\rho_{10}=\rho_{12}$). 
\end{itemize}
On the other hand, Fig. S3(b) shows that $\Delta\left\langle \mathcal{F}_{7,\delta}\right\rangle <0$
for $2.3\lesssim\delta\lesssim3.8$. This implies that GP can provide
detection of a hidden environment in a situation where tests based
on the RT constraints fail to detect the heat leak. Having said that,
it is important to mention that global passivity refers to constraints
on the total system, and not just on a subsystem as in the case of
resource theory. If the total system is very large resource theory
could have the advantage of requiring to evaluate its inequalities
only on a small subsystem. On the other hand, we also note that in
contrast to the global passivity constraints, the free energies of
resource theory are not observables in the sense of representing mean
values of hermitian operators. In addition, the violation of a RT
inequality can reliably diagnose the presence of the environment as
long as the evolution without the environment is energy-preserving.
Otherwise, a violation of the RT inequalities may indicate the exchange
of work, and not necessarily the existent of a heat leak.  \bibliographystyle{apsrev}
\bibliography{Refs}

\end{document}